\documentclass[aps,prb,twocolumn,showpacs,unsortedaddress,amsmath,amssymb,10pt, floatfix]{revtex4-1}
\usepackage[utf8x]{inputenc}
\usepackage{amsmath}
\usepackage{amsfonts}
\usepackage{bbm}               % for getting a nice 1
\usepackage[hyperfootnotes=false,pdfpagelabels,pagebackref,unicode]{hyperref}
\usepackage[caption=false]{subfig}
\usepackage{relsize}
\usepackage{placeins}
\usepackage{pstool}
\usepackage{bbold}
\usepackage{ulem}
\usepackage{xstring}
\EndPreamble

\makeatletter
\def\tagform@#1{\maketag@@@{\ignorespaces#1\unskip\@@italiccorr}}
\let\orgtheequation\theequation
\def\theequation{(\orgtheequation)}
\makeatother
\newsavebox{\mytmpbox}

\newcommand{\img}[3]%
{{
\immediate\write18{cp ./gfx/#2.eps ./#2.eps}%
\sbox{\mytmpbox}{\psfragfig*[#1]{#2}{%create the pdf
#3}}
\immediate\write18{mv #2.pdf gfx/}
\immediate\write18{rm #2.eps}
\includegraphics[#1]{#2}%}
}}

\let\orgautoref\autoref
\renewcommand{\autoref}%
        {%
         \orgautoref}

\begin{document}
\bibliographystyle{utcaps}
\title{Equivalence of Rashba-Hubbard and Hubbard chains}
\author{Florian Goth}
\author{Fakher F. Assaad}
\affiliation{Institut f\"ur Theoretische Physik und Astrophysik,\\
Universit\"at W\"urzburg, Am Hubland, D-97074 W\"urzburg, Germany}
\date{\today}

\begin{abstract}
We review the fact that $U(1)$ gauge symmetry enables the mapping of one-dimensional Hubbard chains 
with Rashba-type spin orbit coupling to renormalized Hubbard Hamiltonians.
The existence of the mapping has important consequences for the interpretation of ARPES experiments
on one-dimensional chains subject to Rashba spin orbit coupling
and can be exploited to check for the applicability of the mapping.
We show numerical applications of the mapping
and consider the implications for bosonization as well as for the Heisenberg-limit of the Hubbard model.
In addition we point out the consequences for various generalized Hubbard models.
\end{abstract}
\maketitle
\section{Introduction}
The steady advances in controlling the surface of a solid coupled 
with the ingenuity of experimental groups\cite{PhysRevLett.111.137203, PhysRevB.82.161410, 1367-2630-15-10-105013,0953-8984-25-1-014013}
have made it possible to grow monatomic chains in a controlled manner.
Since in this setup inversion symmetry is broken
the Rashba spin orbit interaction is not negligible 
and its coupling strength is proportional to the gradient of the electrical potential
perpendicular to the surface.
Since these chains are now within reach of our experimental devices 
the question is now asked to theory which behaviour is expected in this 
setup with strong spin-orbit coupling.
A very detailed answer to this problem has already been given a long time
ago by Kaplan \cite{Kaplan1983}.
He gave a detailed description how the single band Hubbard model
with Rashba spin orbit interaction is solved in terms of the same model without 
spin orbit interaction by a gauge transform of the fields.
He analytically proved the shift in the single particle spectra and, inspired by work of
Calvo \cite{Calvo1981}, he showed how the spin-spin correlations change in the presence of
spin orbit interaction by considering the strong coupling limit of the Hubbard model,
the Heisenberg chain.
Some years later the same transform was found by Meir et al. and used to study disorder in mesoscopic rings in Ref.~ \onlinecite{PhysRevLett.63.798}.
Persistent currents in mesoscopic Hubbard rings with this particular form of spin orbit interaction
were studied by Fujimoto et al in Ref.~\onlinecite{PhysRevB.48.17406}.
To our knowledge they were the first that noted the interpretation in terms of a comoving frame of reference 
for the spin quantization axis. This peculiar rotation was also noted by Refs.~\onlinecite{Kaplan1983,PhysRevLett.97.236601,PhysRevB.82.045127}.
The transformation is naturally present in one dimension but also exists in higher
dimension as already pointed out by Kaplan\cite{Kaplan1983}.
The realization is possible if one considers the interplay of Rashba and Dresselhaus
spin orbit interaction at special values of the coupling strength\cite{PhysRevLett.90.146801,PhysRevLett.97.236601}
which has been experimentally realized in ultracold atom experiments\cite{PhysRevLett.109.095301,PhysRevLett.109.095302}.
Another possibility is to assume the existence of a vector potential with a specific direction\cite{PhysRevB.82.161305,PhysRevA.87.041602}
that generates the spin orbit interaction.
Aspects of this mapping in a bosonization context were already mentioned in Ref.~\onlinecite{PhysRevB.82.045127} which discussed the special case of an infinite parabolic band
and considered the Peierls transition in 1D systems.
Results on the critical exponents for this model using Bethe Ansatz were
obtained by Zyvagin\cite{PhysRevB.86.085126}. 
In this paper we review the fact that despite the complications due to the seemingly more complex band structure
because of the spin orbit interaction all results can be connected back to the familiar Hubbard model.
Since experimental groups now have the possibility to study these surface chains
we focus on the details of finite lattices with finite bandwidth at non-zero temperature and its experimental consequences,
which should also be of importance to the field of ultracold fermionic chains\cite{PhysRevB.88.165101,PhysRevB.88.205127}.
The importance of this mapping has grown since algorithmic progress allows for a very precise numerical evaluation
of spectral and thermodynamic properties of the Hubbard model.
The mapping now enables them to address
systems with Rashba spin-orbit interaction using almost the same codes.
In this paper we will reinterpret simulations of the Hubbard model in the Rashba-Hubbard setting.
As a further consequence of the mapping we will consider the strong coupling Heisenberg limit for the half-filled band and give spectra 
for spin spin correlations.
The structure of the paper is as follows:
We will first pin down the lattice Hamiltonian model in \autoref{sec:model}.
in \autoref{sec:identity} we derive in detail the mapping from the Rashba-Hubbard chain 
to the plain Hubbard model at the Hamiltonian level. This section is finished 
by \autoref{subsec:observables} which discusses the  consequences for observables.
This sets the stage for \autoref{sec:consequences} which in particular
discusses the experimental consequences for spin resolved ARPES 
measurements. Here we discuss experimental tests on the vaildity of the mapping.
In \autoref{sec:limits} we study the consequences
for limiting cases of the Hubbard model, namely for a bosonization treatment
and for the Heisenberg limit. We will compare correlation 
functions derived from both limits to actual Monte-Carlo data.
After that we briefly consider some generalizations in \autoref{sec:generalizations}
before we conclude the paper with \autoref{sec:summary} and give an outlook.

\section{The model and its Hamiltonian}
\label{sec:model}
We consider the Hamiltonian
\begin{equation}
 H = H_t + H_U + H_\lambda
 \label{eq:H_basic}
\end{equation}
with the bare hopping Hamiltonian
\begin{equation}
\begin{split}
  H_t & = -t \sum \limits_{r} \vec{c}^{\hspace{0.1em}\dagger}_{r} \vec{c}^{\phantom{\dagger}}_{r+1} + h.c. - \mu N \\
  &= -2t \sum \limits_{k\sigma} \cos(k) n_{k,\sigma} -\mu N\\
\end{split}
\end{equation}
of electrons on a linear chain.
$\vec{c}_r$ denotes a spinor of fermionic operators $c^\dagger_{r,\sigma}$ ($c^{\phantom{\dagger}}_{r,\sigma}$)
which create (annihilate) an electron at site $r$ with spin $\sigma$.
Here $t$ denotes the hopping matrix element which is set to $t=1$ for all that follows and $n_{k,\sigma} = c^\dagger_{r, \sigma} c^{\phantom{\dagger}}_{r, \sigma}$.
The Rashba-type spin orbit interaction(SOI) reads
\begin{equation}
\begin{split}
  H_\lambda &= \lambda \sum \limits_r \vec{c}^{\hspace{0.1em}\dagger}_{r+1} i \sigma^{\phantom{\dagger}}_y \vec{c}^{\phantom{\dagger}}_r + h.c. \\
  &= -2 \lambda \sum \limits_{k,\sigma} \sigma i \sin(k) c^\dagger_{k,-\sigma} c^{\phantom{\dagger}}_{k,\sigma}.
\end{split}
\end{equation}
The Rashba spin orbit coupling with coupling strength $\lambda$ breaks $SU(2)$ spin symmetry but preserves 
time-reversal symmetry. $\sigma_i$ with $i \in \{x,y,z\}$ denotes the set of three Pauli spin matrices.
Finally the Hubbard interaction is written in a particle-hole symmetric form:
\begin{equation}
  H_U  = U \sum \limits_r \left(n_{r,\uparrow} - \frac{1}{2} \right) \left(n_{r,\downarrow} - \frac{1}{2} \right).
  \label{eq:Hubbard_Interaction}
\end{equation}
Its strength is given by $U$.
Note that in all these equations $\sigma \in \{\uparrow, \downarrow \}$ denotes the physical spin and periodic boundary conditions are imposed.
\section{The Hamiltonian identity}
\label{sec:identity}
\subsection{The helical base}
First we diagonalize the non-interacting part given by 
\begin{equation}
\begin{split}
 H_0 & = H_t + H_\lambda\\
 &= \sum_{k\sigma} \epsilon(k) c^\dagger_{k,\sigma} c^{\phantom{\dagger}}_{k,\sigma} +  i \sigma V(k) c^\dagger_{k,-\sigma} c^{\phantom{\dagger}}_{k,\sigma} - \mu N
\end{split}
\end{equation}
with $\epsilon(k) = -2 \cos(k)$ and $V(k) = -2\lambda \sin(k)$.
The Hamiltonian is already diagonal in $k$-space therefore we only need to perform a rotation in spin-space to fully diagonalize it.
A possible rotation around the x-axis is given by
\begin{equation}
 S = \frac{1}{\sqrt{2}}(1 + i \sigma_x),
 \label{eq:fermionic_transformation}
\end{equation}
such that
\begin{equation}
 c_{k,s} = \sum_\sigma S_{s,\sigma} c_{k,\sigma}.
\end{equation}
The index $s \ \in \{+,-\}$ will exclusively refer to the new "helical" fermions given by the above equation.
The effect is that we have rotated $\sigma_y$ onto $\sigma_z$.
Note that \eqref{eq:fermionic_transformation} is valid irrespective of the internal structure of $\epsilon(k)$ and $V(k)$.
Performing the algebra the non-interacting Hamiltonian then reads
\begin{equation}
 H_0 = -2 \sum \limits_{k,s} E^s(k) n_{k,s} - \mu N
 \label{eq:H0_helical}
\end{equation}
with the new non-interacting dispersion given by
\begin{equation}
 E^s(k) = \cos(k) - \lambda s \sin(k).
\end{equation}
Using the harmonic addition theorem this can be recast into the form
\begin{equation}
 E^s(k) = \sqrt{1+\lambda^2} \cos(k - s \phi(\lambda))
 \label{eq:disp0}
\end{equation}
where $\phi(\lambda) = \arctan(\lambda)$. Simple as this step may seem it is enlightening since
it shows that the effect of the Rashba-interaction can be separated into an increase of the bandwidth given
by $\sqrt{1 + \lambda^2}$ and a phase-shift $\phi(\lambda)$ that differs in sign for the different helicity
branches.
In terms of helical fermions we find from \eqref{eq:fermionic_transformation} for the spin resolved particle densities:
\begin{equation}
 n_{k,\sigma} = \frac{1}{2} \left(  n_{k,+} + n_{k,-} - \sigma c^\dagger_{k,+} c^{\phantom{\dagger}}_{k,-} - \sigma c^\dagger_{k,-} c^{\phantom{\dagger}}_{k,+}\right).
\end{equation}
Inserting this into the Hubbard Interaction \eqref{eq:Hubbard_Interaction} we see that it stays form-invariant under this transformation. We find 
\begin{equation}
  H_U  = U \sum \limits_r \left(n_{r,+} - \frac{1}{2} \right) \left(n_{r,-} - \frac{1}{2} \right)
  \label{eq:Hubbard_Interaction_helical}
\end{equation}
with the helical particle densities $n_{r, s}$.
\subsection{The phase-shift and gauge symmetry}
The phase-shift is given by the Arc Tangent, $\phi(\lambda) = \arctan(\lambda)$.
The following addition theorem holds for $\arctan(x)$:
\begin{equation}
 \arctan(x) + \arctan(y) = \arctan\left( \frac{x + y}{1-xy}\right) +\frac{\pi}{2} g(x,y)
 \label{eq:additiontheoremarctan}
\end{equation}
with $x y \neq 1$ and 
\begin{equation}
 g(x,y) = \text{sgn}(x)(\text{sgn}(x y - 1)+1)
\end{equation}
Setting $x = \lambda$ we now impose the condition
\begin{equation}
 \arctan\left( \frac{\lambda + y}{1-\lambda y}\right) = \frac{j \pi}{m}
 \label{eq:xlinky}
\end{equation}
where $m>2, 1 \leqq j < m $ denote integers with a greatest common divisor of one.
Defining 
\begin{equation}
t_m = \tan(\frac{j\pi}{m}) = \frac{s_m}{c_m} = \frac{\sin(\frac{j\pi}{m})}{\cos(\frac{j\pi}{m})},
\end{equation}
\eqref{eq:xlinky} links $\lambda$ and $y$ in the following way:
\begin{equation}
 y = \frac{t_m - \lambda}{1 + t_m \lambda}.
\end{equation}
Next we check the requirement that $\lambda y \neq 1$:
\begin{equation}
\begin{split}
 \lambda y & = \frac{\lambda(t_m - \lambda)}{1 + t_m \lambda} \neq 1 \\
 \rightarrow & \lambda t_m - \lambda^2 \neq 1 + t_m \lambda \\
 \rightarrow & \lambda^2 \neq -1.\\
 \end{split}
\end{equation}
So this should hold for all real $\lambda$.
Next we consider $g(\lambda, y)$:
 \begin{equation}
\begin{split}
 g(\lambda, y) & = \text{sgn}(\lambda)(\text{sgn}(\frac{\lambda (t_m - \lambda)}{1 + t_m \lambda} - 1) + 1)\\
 & = \text{sgn}(\lambda)(\text{sgn}(\frac{-1 - \lambda^2}{1 + t_m \lambda}) + 1) =: \sigma_m(\lambda)\\
\end{split}
\end{equation}
We note that $\sigma_m(\lambda)$ can only take the values $0$ and $\pm 2$.
Combining all those preliminaries we find the following identities parametrized by $m$ for the phase-shift
\begin{equation}
 \phi(\lambda) = \frac{j\pi}{m}+ \frac{\pi}{2}\sigma_m(\lambda)+\phi(\lambda_m)
 \label{eq:phi-identity}
\end{equation}
with the definition of the new SOI
\begin{equation}
\begin{split}
 \lambda_m &= \frac{\lambda - t_m}{1 + t_m \lambda} \\
 &= \frac{c_m \lambda -s_m}{s_m \lambda + c_m}.
\end{split}
\label{eq:lambda_new}
\end{equation}
Note that \eqref{eq:lambda_new} is an elliptic M\"obius transform.
\eqref{eq:lambda_new} forms a cyclic group of length $m$ which is the reason that the $j$ dependence 
is suppressed since the transform for $j=1$ is the generator for all elements by repeated insertion.
Now we need to apply this to the non-interacting Hamiltonian. It is convenient to transform the non-interacting Hamiltonian given in the helical base in 
\eqref{eq:H0_helical} back to real-space. We find 
\begin{equation}
 H_0(\lambda) = -\sqrt{1+\lambda^2} \sum \limits_{rs} e^{i s \phi(\lambda)} c^\dagger_{r+1, s} c_{r,s} + h.c.
\end{equation}
and see that in real space the effect of the phase-shift is that of a helicity dependent magnetic flux. Inserting \eqref{eq:phi-identity}
and employing gauge invariance we find
\begin{equation}
 \begin{split}
  &H_0(\lambda) \\
  &=-\sqrt{1 + \lambda^2} \sum_{rs} e^{i s \left( \frac{j\pi}{m} + \frac{\pi}{2}\sigma_m(\lambda) + \phi(\lambda_m)\right)}c^\dagger_{r+1,s} c^{\phantom{\dagger}}_{r,s} + h.c.\\
%  &=-\sqrt{1 + \lambda^2} \sum_{rs} e^{i s \left( \frac{j\pi}{m} + \frac{\pi}{2}\sigma_m(\lambda)\right)}  e^{i s\phi(\lambda_m)}c^\dagger_{r+1,s} c_{rs} + h.c.\\
  & = -\nu_m(\lambda) \sqrt{1 + \lambda_m^2} \sum_{rs} e^{i s \phi(\lambda_m)} \tilde{c}^\dagger_{r+1,s} \tilde{c}^{\phantom{\dagger}}_{r,s} + h.c.\\
  & = \nu_m(\lambda) H_0(\lambda_m).
  \label{eq:Ham0}
 \end{split}
\end{equation}
We have defined the scaling-factor
\begin{equation}
\begin{split}
 \nu_m(\lambda) &= \sqrt{\frac{1 + \lambda^2}{1 + \lambda_m^2}}\\
 &= |\cos(\frac{j\pi}{m}) + \sin(\frac{j\pi}{m}) \lambda|
\end{split}
\end{equation}
as well as gauge-transformed new fermionic operators
\begin{equation}
 \tilde{c}_{r,s} = c_{r,s} e^{-i s r (\frac{j\pi}{m} + \frac{\pi}{2}\sigma_m(\lambda))}
 \label{eq:gen_gauge_transform}
\end{equation}
An important ingredient is how the boundary conditions transform under this choice of the gauge.
Our original operators were subject to the condition $c_{r+L, s} = c_{r,s}$. This is fulfilled for $\tilde{c}_{r,s}$, if 
\begin{equation}
 \tilde{c}_{r+L,s}e^{-i s (r+L) (\frac{j\pi}{m} + \frac{\pi}{2}\sigma_m(\lambda))} = c_{r,s} e^{-i s r (\frac{j\pi}{m} + \frac{\pi}{2}\sigma_m(\lambda))}
\end{equation}
and therefore
\begin{equation}
 e^{-i s L (\frac{j\pi}{m} + \frac{\pi}{2}\sigma_m(\lambda))} = 1
\end{equation}
which is equivalent to
\begin{equation}
  s L (\frac{j\pi}{m} + \frac{\pi}{2}\sigma_m(\lambda)) = 2\pi n
\end{equation}
with an arbitrary integer $n$. Simplifying we find
\begin{equation}
 L = \frac{4 n m s}{2 j + m \sigma_m(\lambda)}
\label{eq:boundary}
\end{equation}
In total we now have for the Hamiltonian the following identity:
\begin{equation}
H(\lambda, \mu, U) = \nu_m(\lambda) H(\lambda_m, \frac{\mu}{\nu_m(\lambda)}, \frac{U}{\nu_m(\lambda)})
\end{equation}
This means that for a given $\lambda$, $H(\lambda, \mu, U)$
is connected to $m$ other Hamiltonians that are identical at the Hamiltonian level,
with the new SOI given by $\lambda_m$.
Now we want one of those points to be the plain Hubbard model at $\lambda=0$.
For that we have to start to explore the consequences 
of the equation 
\begin{equation}
 \lambda_{m}(\lambda) = 0
\end{equation}
which implies 
\begin{equation}
 t_{m} = \tan\left( \frac{j\pi}{m} \right) = \lambda.
 \label{eq:newlambdas}
\end{equation}
Or stated in terms of the phase-shift:
\begin{equation}
 \frac{j\pi}{m} = \phi(\lambda).
\end{equation}
This implies $\sigma_m(\lambda) =  0$.
For the scaling factor we find
\begin{equation}
 \nu_{m} = \sqrt{1+ t_{m}^2} = \frac{1}{|c_{m}|}.
\end{equation}
Stated in terms of an operator identity between Hamiltonians we derive from 
\eqref{eq:Ham0}
the relation
\begin{equation}
 H_0(t_{m}) = \sqrt{1 + t_{m}^2} H_0(0) = \frac{H_0(0)}{|c_{m}|}.
\label{eq:transformH0}
\end{equation}
Which means for the full Hamiltonian \eqref{eq:H_basic}
\begin{equation}
 H(t_m, \mu, U) = \frac{H(0, \mu |c_{m}|, U |c_{m}|)}{|c_{m}|}.
 \label{eq:scaled_H}
\end{equation}
Of course identities \eqref{eq:transformH0} and \eqref{eq:scaled_H} are independent on the choice of the ensemble used for the thermal averaging.
Hence the mapping is valid in the canonical as well as in the grand-canonical ensemble.
Setting $m=L$ and compatibility with \eqref{eq:boundary} requires $j=2n$.
This can be interpreted as follows: For a given system size $L$ the plain Hubbard model is connected to $\frac{L-1}{2}$ points on the real $\lambda$ axis corresponding to a 
Hamiltonian with a spin orbit strength determined by
$\lambda = \tan(\frac{2j \pi}{L})$.
In the limit  $L\rightarrow \infty$ $\lambda$ therefore becomes continuous.
Note that these are precisely the points where $\tan(x)$ takes algebraic values.
In \autoref{fig:spectra} we show a selection of spectra which are connected to a single simulation of the plain Hubbard model via the identity
\eqref{eq:scaled_H}.
To conclude this section we give a geometric perspective on the group structure in \eqref{eq:lambda_new}.
\eqref{eq:lambda_new} exhibits a cyclic group structure of length $n$ which is the symmetry group of the regular
$n$-gon which is also the symmetry group of the roots of
\begin{equation}
z^{n} = 1,
\label{eq:ngon}
\end{equation}
the roots of unity, in the complex plane.
For a finite chain with arbitrary $\lambda$ it is not granted that there exists a root that is located at $z_0=1$. The $n$-gon will be slightly canted with respect to the solutions of \eqref{eq:ngon}
and is given by the solutions of 
\begin{equation}
 z^n = e^{i \alpha n}
\label{eq:cantedngon}
\end{equation}
with some arbitrary angle in the complex plane $\alpha$.
The freedom to rotate the $n$-gon from one root to the next is given
by the $U(1)$ gauge symmetry of quantum mechanics.
The restriction on the values of $\lambda$ in \eqref{eq:newlambdas} now ensures that $\alpha=0$ in \eqref{eq:cantedngon} and therefore that exactly one of the roots is located at $z_0=1$ which corresponds to the 
plain Hubbard model and an angle of rotation between the roots of $\phi(\lambda) = \arctan(\lambda)$.
As seen also by Ref.~\onlinecite{PhysRevLett.97.236601} for a 2D system
this model exhibits an $SU(2)$ symmetry since we have mapped it to the plain Hubbard model.
Of course this is subject to a proper transformation of the boundary conditions.
\subsection{Consequences for observables}
\label{subsec:observables}
The rescaling due to \eqref{eq:scaled_H} as well as the transformation of the operators in \eqref{eq:gen_gauge_transform}
forces us to transform our physical quantities as well.
In $k$-space we have for the fermionic operators
\begin{equation}
 c_{k,s} = \tilde{c}_{k + \frac{s j \pi}{m},s}
\end{equation}
Using this we find for observables in the helical base:
\begin{equation}
\begin{split}
 n_{k,s} &= \tilde{n}_{k + \frac{s j \pi}{m},s},\\
 n_k &= \sum_s n_{k,s} = \sum_s \tilde{n}_{k + \frac{s j \pi}{m},s},\\
 S^z &= \sum_s s n_{k,s} = \sum_s s \tilde{n}_{k + \frac{s j \pi}{m},s},\\
 S^+ &= \tilde{c}^\dagger_{k + \frac{j \pi}{m},+} \tilde{c}^{\phantom{\dagger}}_{k - \frac{j \pi}{m},-},\\
 S^- &= \tilde{c}^\dagger_{k - \frac{j \pi}{m},-} \tilde{c}^{\phantom{\dagger}}_{k + \frac{j \pi}{m},+}.\\
 \end{split}
\end{equation}
Going forward to thermal averages we find
for the single particle Green's function
\begin{equation}
\begin{split}
 &G^s(k,\tau,\beta,\lambda=t_{j,m}, \mu, U) \\
 &= \langle c^\dagger_{k,s}(\tau) c^{\phantom{\dagger}}_{k,s}(0)\rangle \\
% &= \langle \tilde{c}^\dagger_{k+ \frac{s j \pi}{m},s}(\tau) \tilde{c}^{\phantom{\dagger}}_{k+ \frac{s j \pi}{m},s}(0)\rangle \\
 &= \text{Tr}\left( e^{-\beta H(t_{j,m}, \mu, U)} \tilde{c}^\dagger_{k+ \frac{s j \pi}{m},s}(\tau) \tilde{c}^{\phantom{\dagger}}_{k+ \frac{s j \pi}{m},s}(0) \right)\\
 &= \text{Tr}\left( e^{-\tilde{\beta}H(0, \tilde{\mu}, \tilde{U})} \tilde{c}^\dagger_{k+ \frac{s j \pi}{m},s}(\tilde{\tau}) \tilde{c}^{\phantom{\dagger}}_{k+ \frac{s j \pi}{m},s}(0) \right)\\
 &= G^s(k+ \frac{s j \pi}{m},\tilde{\tau},\tilde{\beta},\lambda=0, \tilde{\mu}, \tilde{U} )
\end{split}
\end{equation}
with $\tilde{\tau} = \frac{\tau}{|c_{j,m}|}$, $\tilde{U} =  U |c_{j,m}|$, $\tilde{\mu} = \mu |c_{j,m}|$ and $\tilde{\beta} = \frac{\beta}{|c_{j,m}|}$.
The $j$-dependence on the previously defined quantities $t_m$ and $c_m$ is now written down explicitly since we need to determine the shift in $k$.
Note that the final result on the above equation has $\lambda=0$ and is therefore measured with the plain Hubbard model.
Inverting this equation to explicitly see which point of the Hubbard model is connected to which part of the Rashba-Hubbard chain we have
\begin{equation}
\begin{split}
 &G^s(k_H, \tau_H, \beta_H, \mu_H, U_H) = \\
 &G^s(k - \frac{s j \pi}{m}, \tau_H |c_{j,m}|, \beta_H |c_{j,m}|, t_{j,m}, \frac{\mu}{|c_{j,m}|}, \frac{U}{|c_{j,m}|})
\end{split}
\label{eq:Green_transformed}
\end{equation}
where the index $H$ denotes that the parameter was used in a simulation of the Hubbard model.
\section{Consequences}
\label{sec:consequences}
\subsection{Experimental consequences}
\subsubsection{LDOS}
\begin{figure}
 \begin{tabular}{cc}
 \includegraphics[width=0.48\linewidth]{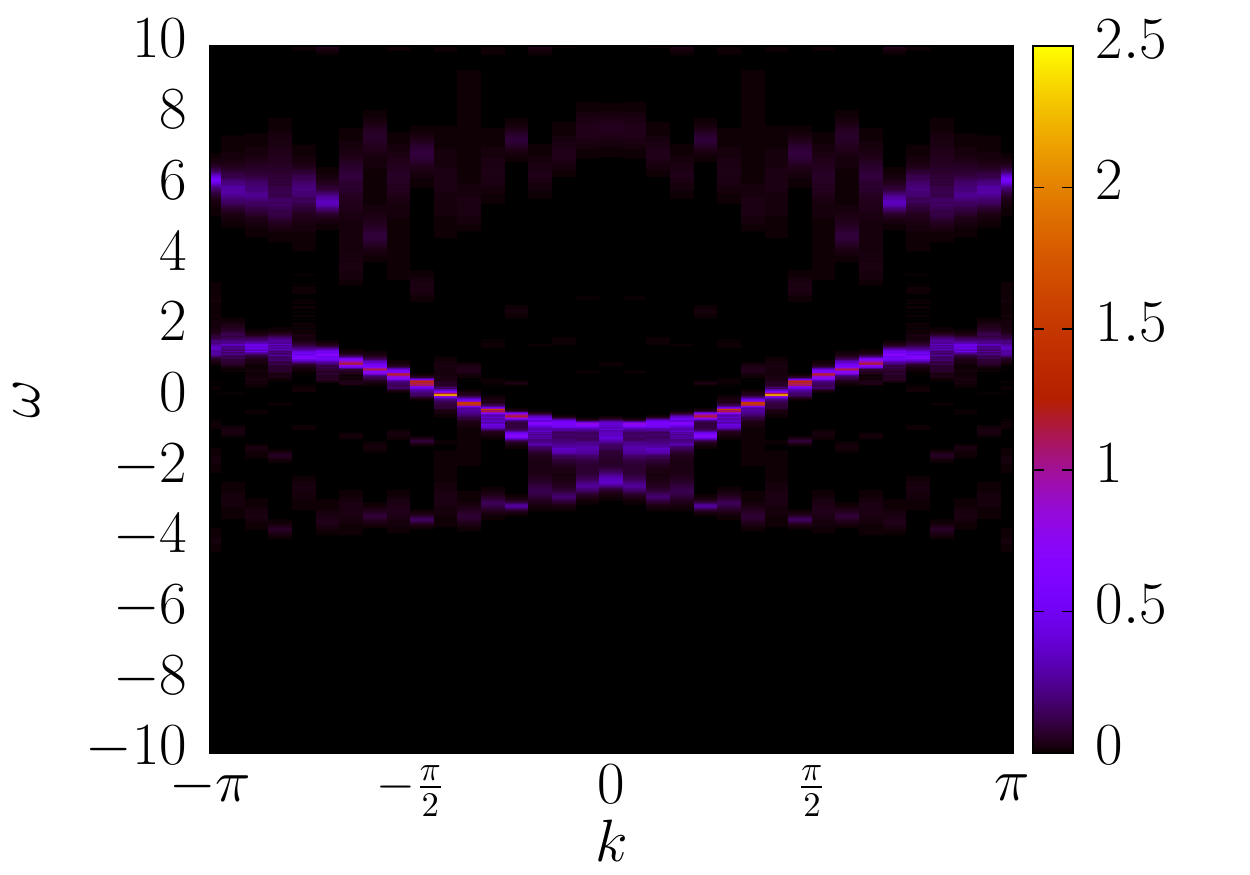} &
 \includegraphics[width=0.48\linewidth]{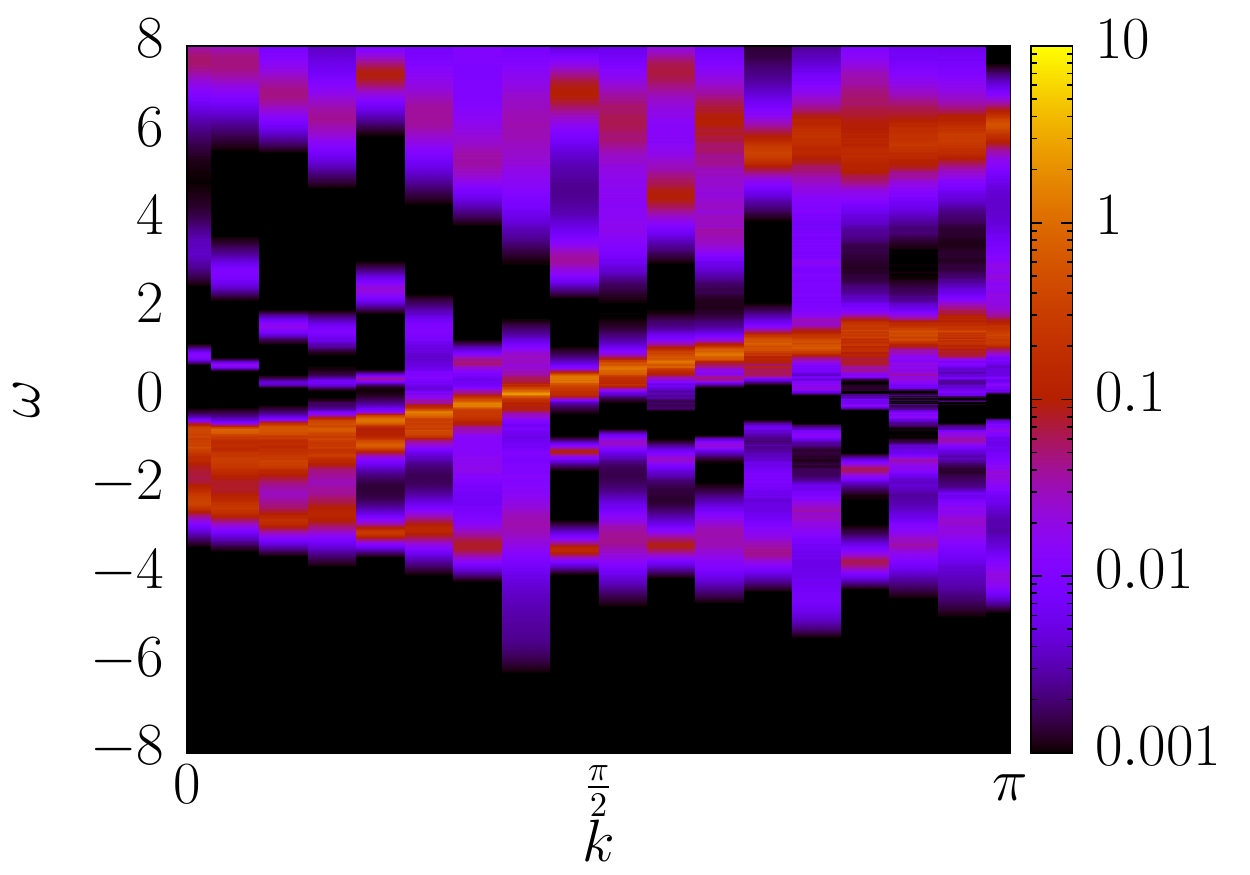} \\
\hspace{-10em}
 \makebox[0pt][l]{(a) $\lambda = 0, U = 6, \beta = 10, \mu = -2.29$} &  \strut\\
 \includegraphics[width=0.48\linewidth]{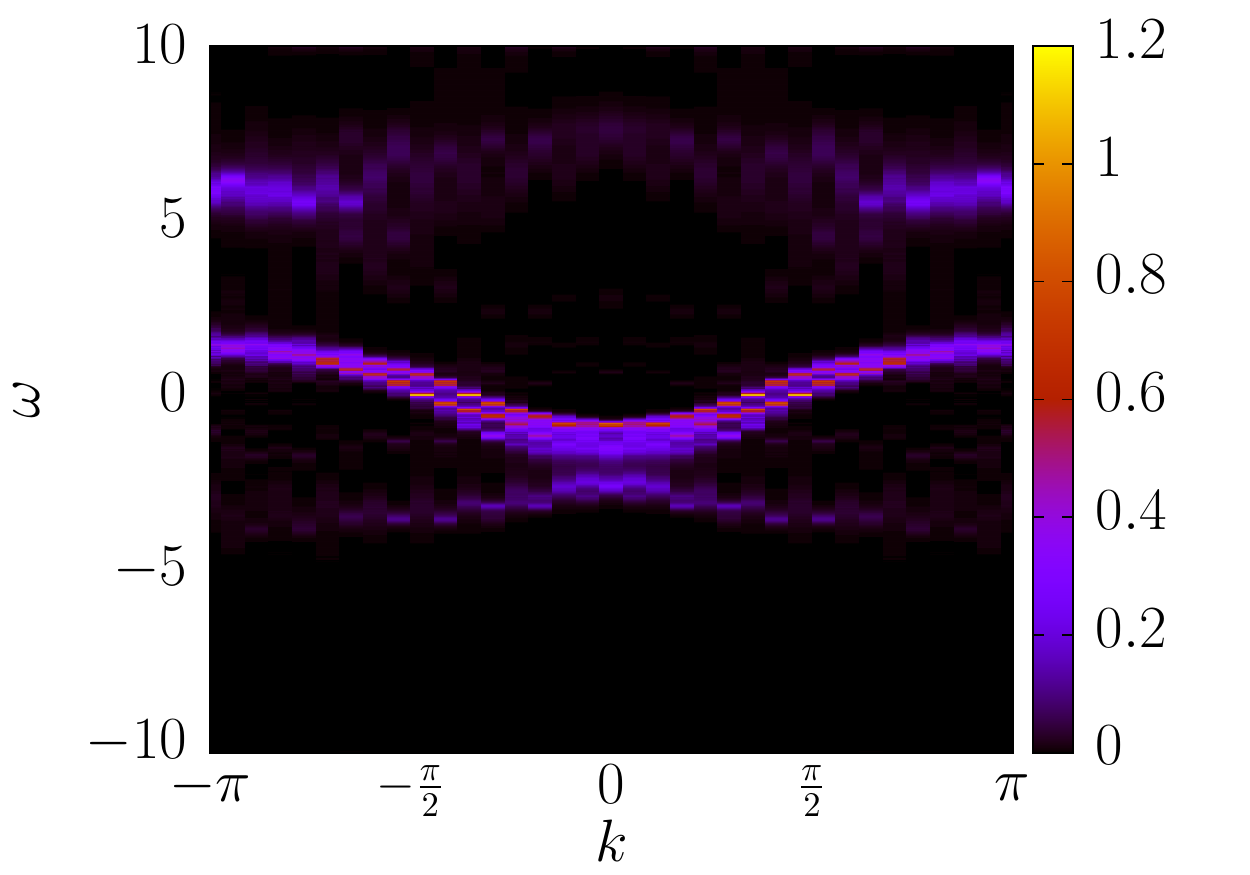} &
 \includegraphics[width=0.48\linewidth]{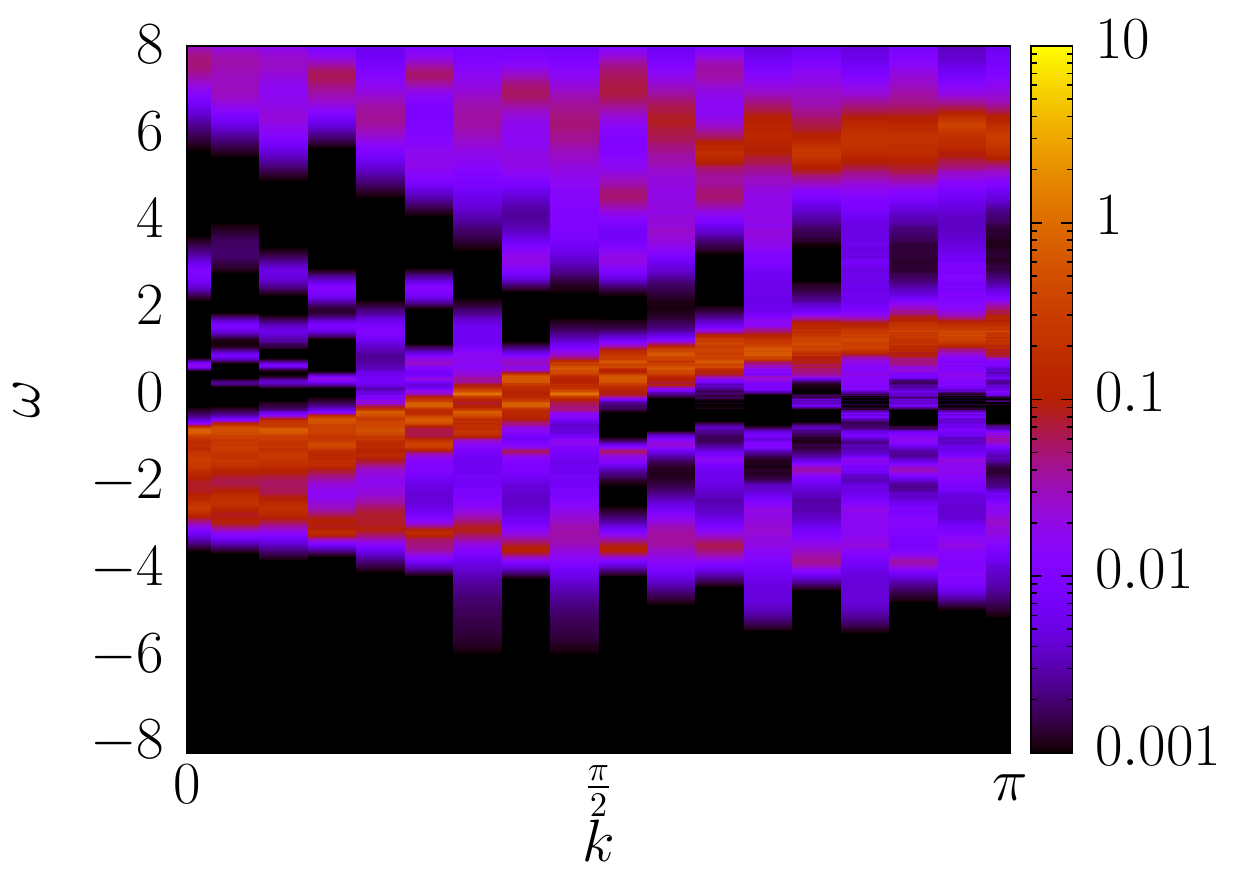} \\
\hspace{-10em}
 \makebox[0pt][l]{(b) $\lambda = 0.186932, U = 6.10, \beta = 9.829730, \mu = -2.32$} &  \strut\\
 \includegraphics[width=0.48\linewidth]{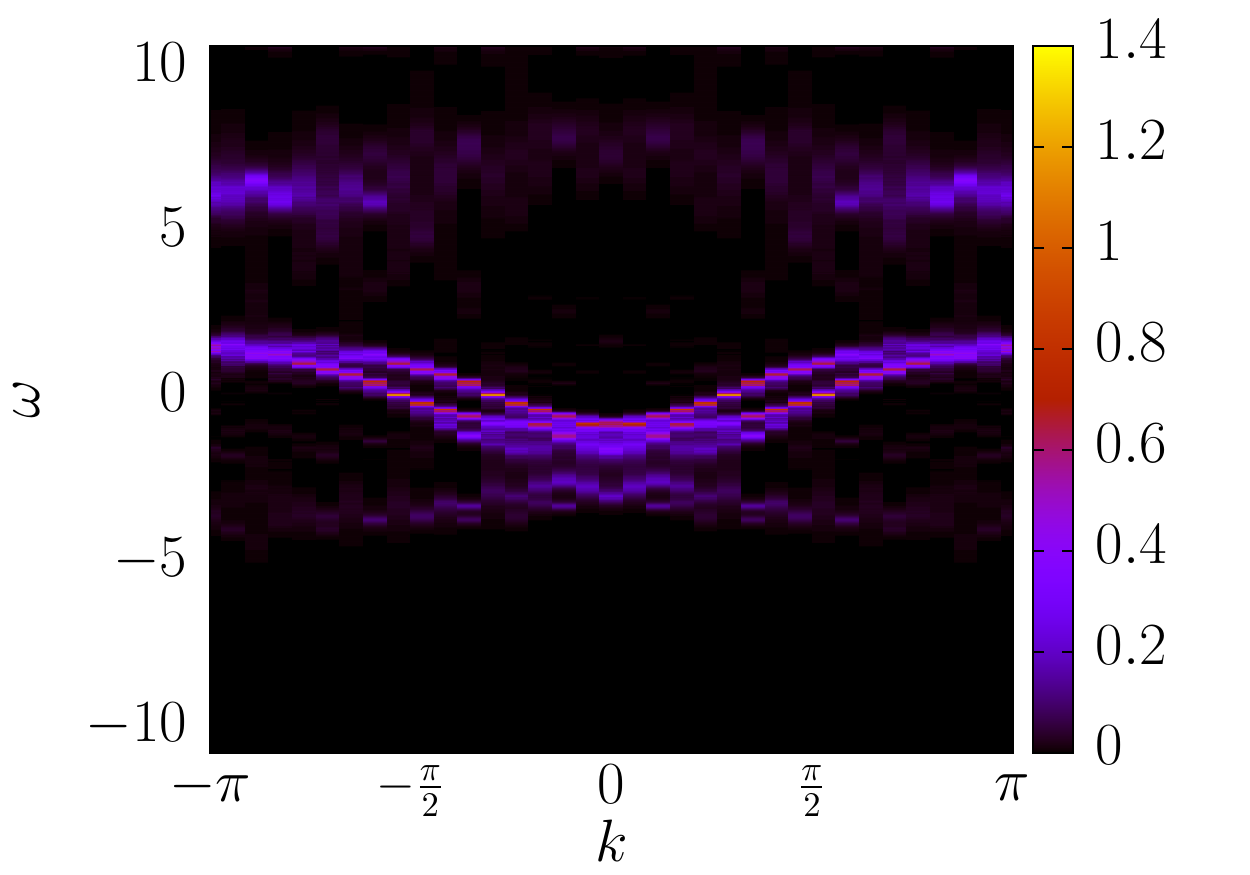} &
 \includegraphics[width=0.48\linewidth]{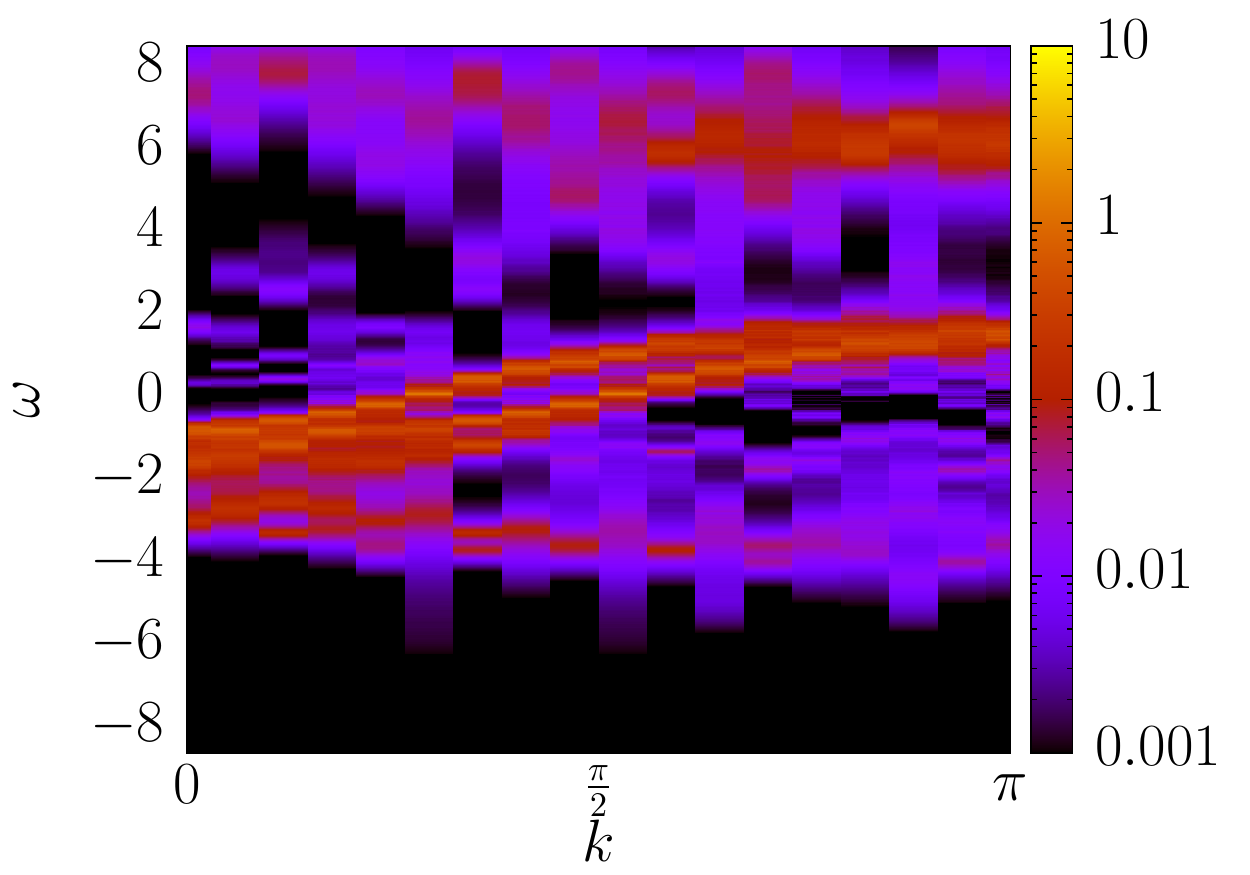} \\
 \hspace{-10em}
 \makebox[0pt][l]{(c) $\lambda = 0.387402, U = 6, \beta = 10, \mu = -2.29$} &  \strut\\
% \includegraphics[width=0.3\linewidth]{Spectra_L34_lambda0_619174_U7_05_B8_502170_mu-2_69.pdf} &\includegraphics[width=0.3\linewidth]{Spectra_L34_lambda0_619174_U7_05_B8_502170_mu-2_69_log_full.pdf} & \includegraphics[width=0.3\linewidth]{Spectra_L34_lambda0_619174_U7_05_B8_502170_mu-2_69_log.pdf} \\
% $\lambda = 0.619174 $ & \strut& \strut\\
 \includegraphics[width=0.48\linewidth]{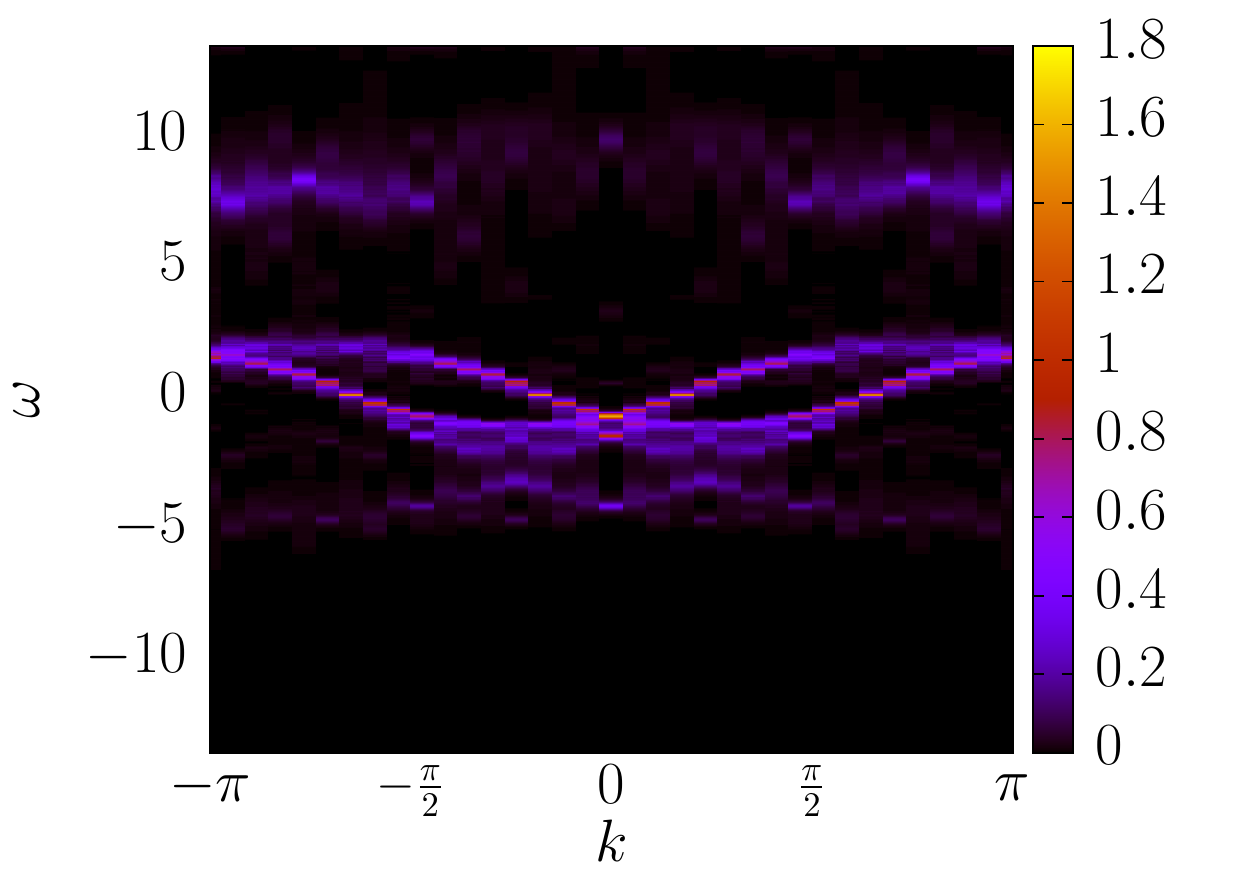} &
 \includegraphics[width=0.48\linewidth]{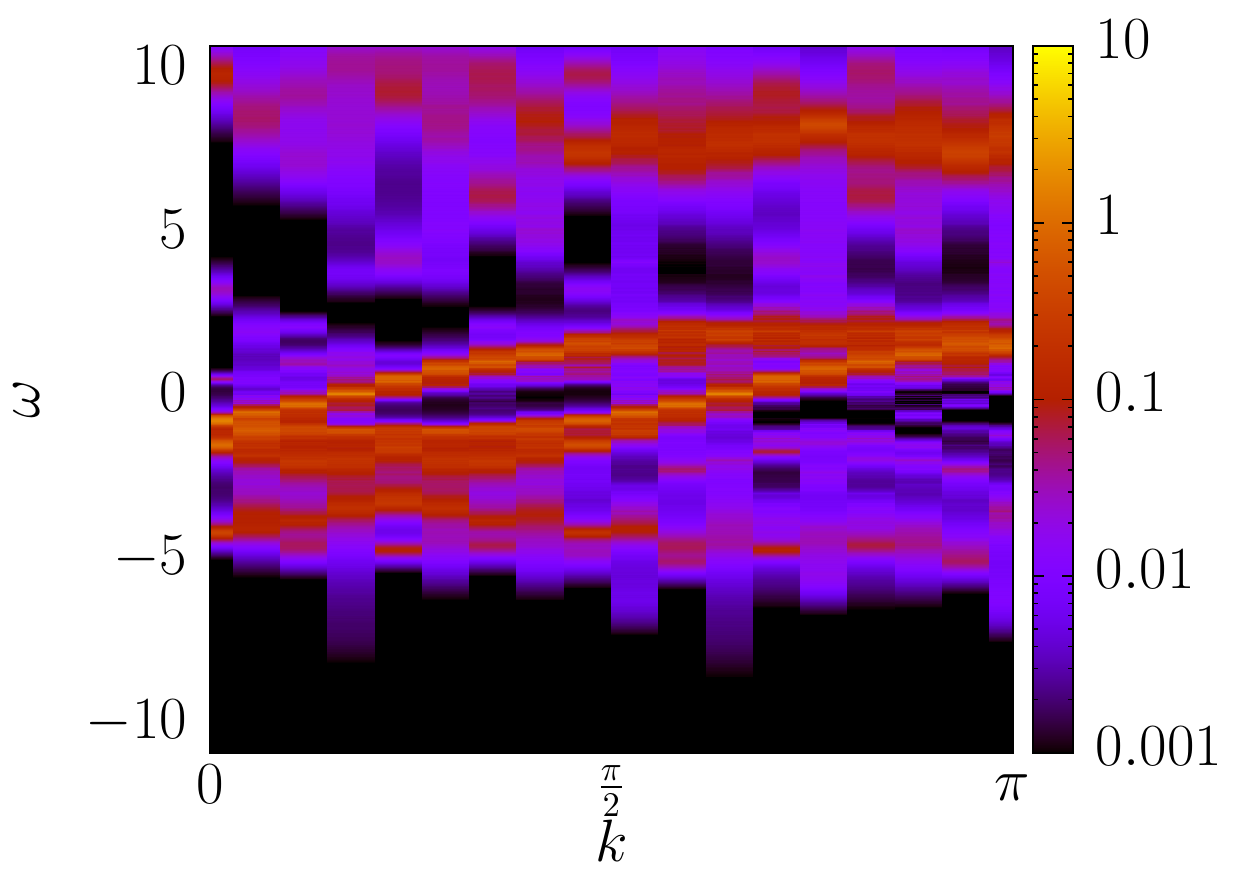}\\
\hspace{-10em}
 \makebox[0pt][l]{(d) $\lambda = 0.91162, U = 8.11, \beta = 7.390090, \mu = -3.09$} &  \strut\\
% \includegraphics[width=0.3\linewidth]{Spectra_L34_lambda1_32421_U9_95_B6_026350_mu-3_79.pdf} &\includegraphics[width=0.3\linewidth]{Spectra_L34_lambda1_32421_U9_95_B6_026350_mu-3_79_log_full.pdf} &\includegraphics[width=0.3\linewidth]{Spectra_L34_lambda1_32421_U9_95_B6_026350_mu-3_79_log.pdf}\\
% $\lambda = 1.32421 $ & \strut& \strut\\
 \includegraphics[width=0.48\linewidth]{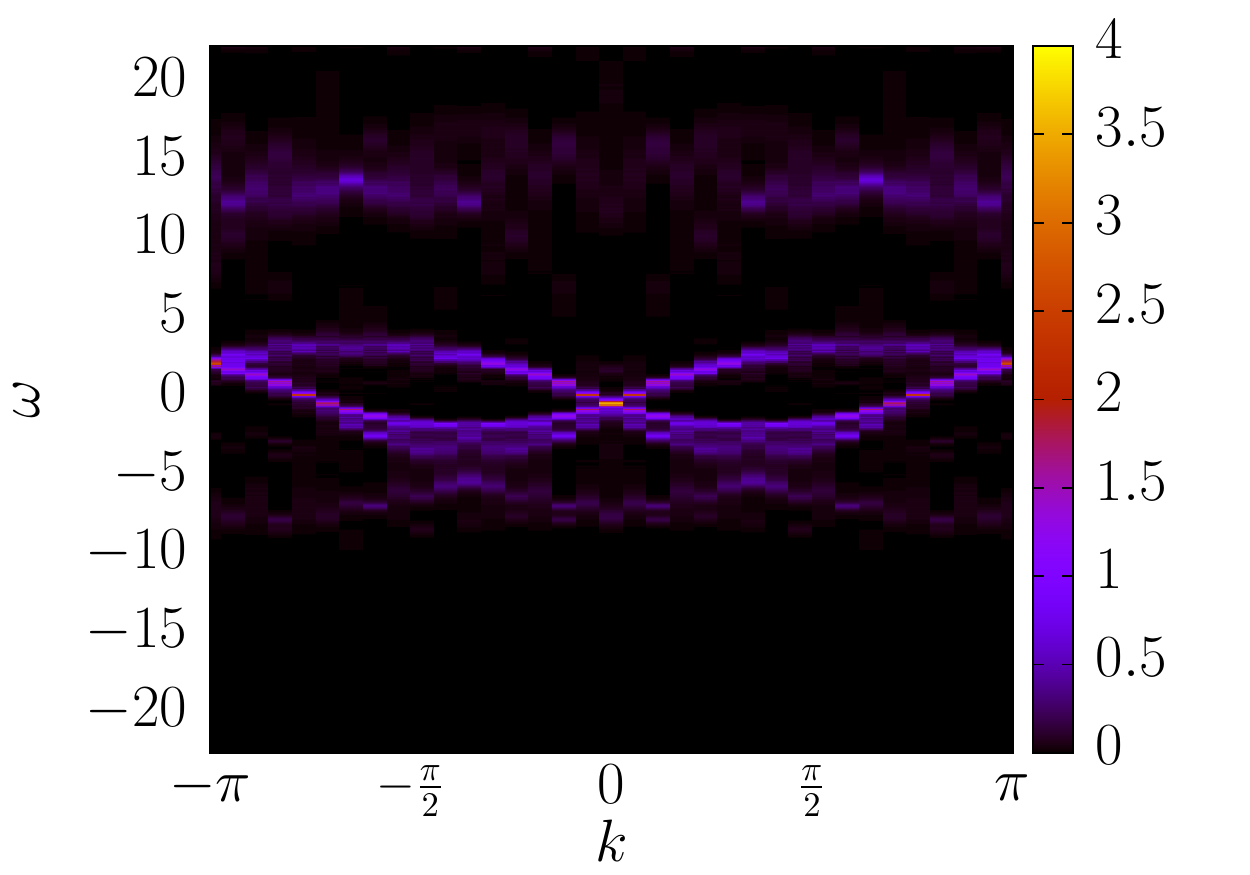} &
 \includegraphics[width=0.48\linewidth]{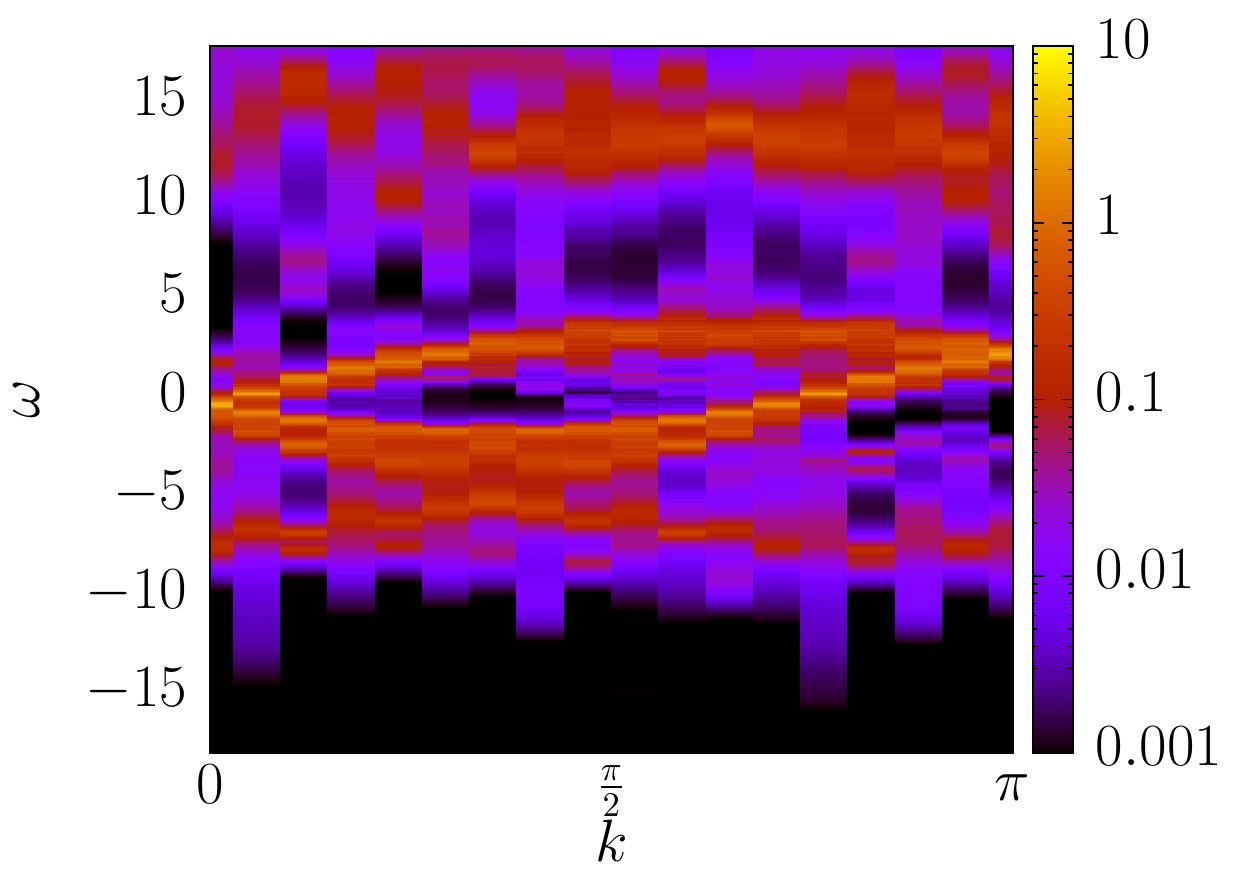}\\
\hspace{-10em}
\makebox[0pt][l]{(e) $\lambda = 2.00827, U = 13.46, \beta = 4.457380, \mu = -5.13$} &  \strut\\
\end{tabular}
\caption{
This panel shows a series of spectra $A(k,\omega)$ which are connected to each other via the mapping \eqref{eq:mapping_spectra}. The starting point is the spectrum in (a) at $\lambda=0$. Increasing $\lambda$ in the panels 
(b) to (d) leads to a shift of the spectra and we see the four Fermi-points developing in the left column. Similarly we see in the logarithmic plots of the right column which are restricted to the domain $[0,\pi]$
the splitting of the original spinon and holon bands.
}
\label{fig:spectra}
\end{figure}

Defining the helicity spin resolved single particle spectral function
\begin{equation}
 A^s(k,\omega) = -\frac{1}{\pi} \text{Im}\left( G^s(k,\omega + i0^+)\right)
\end{equation}
we find that it transforms as
\begin{equation}
 A^s(k,\beta, \omega, \lambda) = |c_{j,m}| A^s\left(k+\frac{s j \pi}{m},\tilde{\beta},\omega |c_{j,m}|, 0\right).
 \label{eq:mapping_spectra}
\end{equation}
%Therefore we find that a quantity like a $k$ and spin summed density of states $D(\omega)$
Hence the local density of states $D(\omega, \beta, \lambda)$ transforms as
\begin{equation}
\begin{split}
 D(\omega, \beta, \lambda) & = \sum_{k,s}  A^s(k,\beta, \omega, \lambda) \\
 & = |c_{j,m}| \sum_{k,s} A^s\left(k+\frac{s j \pi}{m},\tilde{\beta}, \omega|c_{j,m}|, 0\right) \\
 & = |c_{j,m}| D( \omega |c_{j,m}|, \tilde{\beta}, \lambda = 0).
\label{eq:ldos}
 \end{split}
\end{equation}
Therefore the local density of states will not contain any new structure in comparison to the spectra of a plain Hubbard model at the lower temperature  $\tilde{\beta}$.
\autoref{fig:spectra} shows a selection of spectra for various parameters.
The imaginary time Green's functions for \autoref{fig:spectra}a were simulated using an auxiliary field QMC method\cite{Assaad08_rev} along the lines of Ref.~\onlinecite{Abendschein06}
and analytically continued using the stochastic maximum entropy method \cite{Beach04}.
Using \eqref{eq:mapping_spectra} we can then derive the spectra for other values of $\lambda$.
Although these spectra show a seemingly richer structure than the plain Hubbard model,
one can e.g. identify Rashba split holon and spinon bands in the spectra, all spectra have in common that they can be connected back to a single
Hubbard simulation at $U = 6$ and $\beta = 10$ (\autoref{fig:spectra}a) with the well-known signatures of the 
fractionalization of the electron into a spinon and a holon \cite{Abendschein06, Jeckelmann04}.
\subsubsection{Spin resolved spectra}
\begin{figure*}
\subfloat[\strut]{\includegraphics[width=\linewidth]{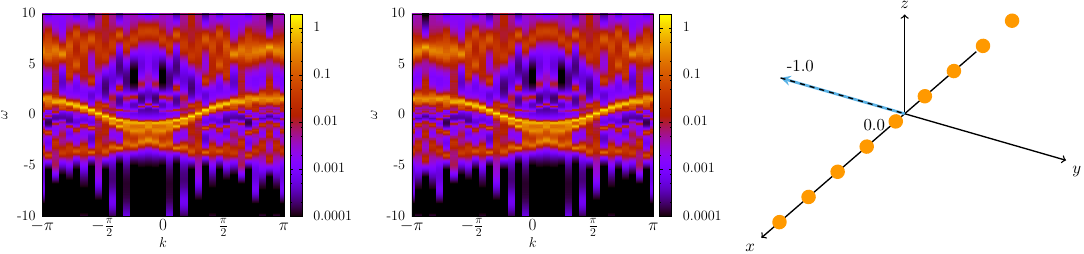}}\\
\subfloat[\strut]{\includegraphics[width=\linewidth]{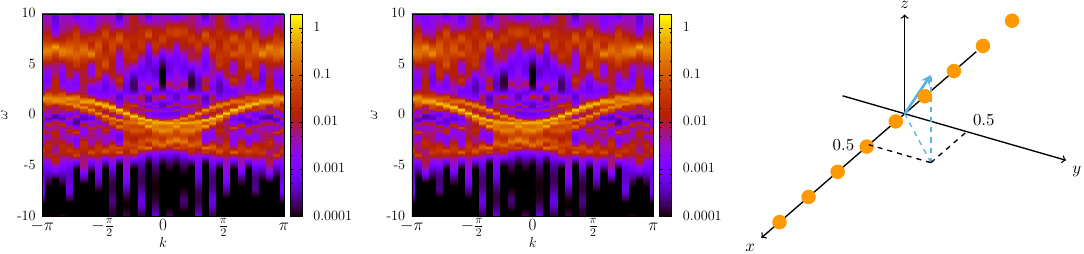}}\\
\subfloat[\strut]{\includegraphics[width=\linewidth]{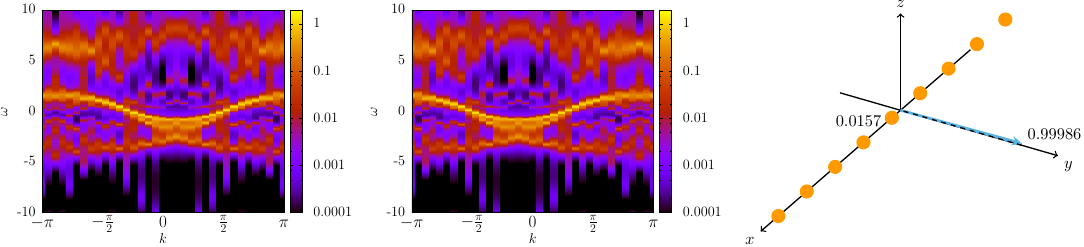}}\\
 \caption{
 Figures showing the different spectra that are obtained using a device with a quantization axis $\vec{D}$ defined by the blue arrow in the right-most column.
 The left-most column shows spectra which have spin up and the spectra in the middle column have spin down with respect to the quantization axis $\vec{D}$.
 The numbers in the right-most column denote the $x$ and $y$ components of $\vec{D}$.
 The spectra are taken from data at $U=6.43, \beta=9.32, \mu=-2.45$ and $\lambda=0.39$ which was in turn derived from a simulation of the Hubbard model at $U=6, \beta=10$ and $\mu=-2.29$.}
 \label{fig:spinspectra}
\end{figure*}

The possibility of doing spin resolved ARPES experiments enables spin resolved measurements of the single particle spectral function.
%If we consider the case of spin-resolved measurements things are a little bit different.
Since the measurement device now defines a preferred spin quantization axis,
we have to calculate the projections of the electrons' original spin quantization axis onto this new axis.
Assuming this axis is given by the usual unit vector $\vec{D} = \left( \sin(\theta) \cos(\varphi), \sin(\theta) \sin(\varphi), \cos(\theta)\right)^T$ with
$\theta \in [0, \pi]$ and $\varphi \in [-\pi, \pi]$ we can rotate
the fermionic operators $\vec{c}_{\vec{e_z}}$, which have $\vec{e_z}$ as quantization axis,
to the new base using
\begin{equation}
 \vec{c}_{\vec{D}} = e^{i \frac{\theta}{2} \vec{\sigma} \cdot \vec{n}} \vec{c}_{\vec{e}_z}.
\end{equation}
Here $\vec{\sigma}$ denotes the Pauli-vector, the rotation axis is $\vec{n} = (-\tan(\varphi), 1, 0)^T$ and 
$\vec{c}_{\vec{D}}$ denotes electrons with the new quantization axis $\vec{D}$.
For Green's functions $G^{\sigma\sigma'}_{\vec{e}_z}$ which have spins measured with respect to the quantization axis $\vec{e}_z$,
and therefore for the spectra, this implies the following relation
\begin{equation}
  G^{\sigma\sigma}_{\vec{D}}(k) = D(k) -\sigma\text{Re}\left( \sin(\theta) e^{-i \varphi} G^{\uparrow \downarrow}_{\vec{e}_z}(k)\right)
\end{equation}
with $D(k) = \sin^2(\frac{\theta}{2}) G^{\uparrow\uparrow}_{\vec{e}_z}(k) + \cos^2(\frac{\theta}{2}) G^{\downarrow\downarrow}_{\vec{e}_z}(k)$.
The spectra shown in \autoref{fig:spectra} correspond to $\vec{D} = \vec{e}_z$. In \autoref{fig:spinspectra}a and \autoref{fig:spinspectra}c we see that along $\vec{D} = \pm\vec{e}_y$
a clear separation of the helicities should be observable, whereas in \autoref{fig:spinspectra}b we see that for a general $\vec{D}$ a mixture of the two helicities is observed.
It is worth noting, that \autoref{fig:spinspectra}a shows that the separation of the helicities is
crystal momentum independent and a signature of the one dimensional nature of the system.
This independence on the variable $k$ should be experimentally observable and, 
only in conjuction with the fact that the spectra in the observed up spin and down spin seperated spectra are equal,
gives a clear indicator whether a plain suitably generalized(see \autoref{sec:generalizations}) Hubbard model is a suitable model system.
\subsection{Numerical consequences}
A lot of codes have been heavily optimized for the solution of Hubbard like problems.
The mapping now enables them to address questions in the Rashba-Hubbard setting.
Since a  direct quantum Monte Carlo(QMC) simulation of \eqref{eq:H_basic} would yield a sign problem,
the existence of the mapping \eqref{eq:scaled_H} is important because it shows that in the 
proper basis (the comoving spin basis) a simulation without the fermionic sign problem is possible
since the plain one dimensional Hubbard model exhibits no sign problem.
Unfortunately we have to trade this fact for a more complicated representation of observables.
Additionally to the single particle Green's function given in \eqref{eq:Green_transformed} we note 
various two-particle correlation functions and their respective equivalents in the Hubbard model analogue.
The left hand side of the equations is measured in a simulation of \eqref{eq:H_basic} and the 
right hand side is measured in the Hubbard model analogue:
\begin{equation}
 \begin{split}
& N(k,\tau, \beta,\lambda, U) = \langle n_{k}(\tau) n_k\rangle \\
& = \sum\limits_{s,s'} \langle n_{k+s\phi, s} (\tilde{\tau}) n_{k+s'\phi,s'}\rangle (\tilde{\beta}, \tilde{U})\\
& S^{zz}(k,\tau, \beta,\lambda, U) = \langle S^z_k(\tau) S^z_k \rangle\\
& = \sum \limits_{s,s'} \langle c^\dagger_{k + s\phi, s}(\tilde{\tau})c^{\phantom{\dagger}}_{k -s\phi, -s}(\tilde{\tau}) c^\dagger_{k + s'\phi, s'}c^{\phantom{\dagger}}_{k -s'\phi, -s'}\rangle(\tilde{\beta}, \tilde{U}).  \\
 \end{split}
\end{equation}
An example of the Green's function is given in \autoref{fig:spectra}, whereas examples
of spin spin correlations are shown in \autoref{fig:szsz} and \autoref{fig:spsm}.
\section{Various limits}
\label{sec:limits}
\subsection{Bosonization}
There are already bosonization studies of the Hamiltonian \eqref{eq:H_basic} in the literature\cite{PhysRevB.62.16900, PhysRevB.84.075466, PhysRevB.82.033407,2013PhRvB..88s5113S}
but since they consider more general setups for the bosonization they have not explicitly written down the connection to the plain Hubbard model.
The connection was mentioned in the context of a bosonization study of Peierls transitions \cite{PhysRevB.82.045127} where the interpretation in terms of a
comoving frame of reference for the spin quantization axis was mentioned.
From \eqref{eq:disp0} we find that the Fermi-wave vectors are given by
\begin{equation}
\begin{split}
k_F^{\alpha,s}(\lambda) &= \alpha \arccos(\frac{-\mu}{2\sqrt{1+\lambda^2}})  + s \phi(\lambda)\\
& = \alpha k_F^0 + s \phi.
\end{split}
\end{equation}
Note that this expression is only well defined if $|\frac{-\mu}{2\sqrt{1+\lambda^2}}| < 1$. This restricts the possible values of $\mu$
to lie within the band. 
Here we have defined the additional index $\alpha \in \{R,L\}$ which enumerates the two possibilities for a band of helicity $s$ to cross the Fermi level.
We also define
\begin{equation}
 k_F^0 = \arccos(\frac{-\mu}{2\sqrt{1+\lambda^2}}).
 \label{eq:kf0}
\end{equation}
Since the Fermi velocity is the group velocity at those two points, we have
\begin{equation}
\begin{split}
 v^\alpha_F(\lambda) & = v_G(k_F^{\alpha,s}(\lambda)) \\
 & = 2 \sqrt{1 + \lambda^2} \sin(\alpha \arccos(\frac{-\mu}{2\sqrt{1+\lambda^2}}) - s \phi + s \phi) \\
 & = \alpha\sqrt{4 (1 + \lambda^2) - \mu^2}.
\end{split}
\label{eq:vf}
\end{equation}
Therefore the absolute value of the Fermi velocity is the same for all helicities it only differs for left and right movers by a different sign, $\alpha$.
\subsubsection{Derivation}
We start the bosonization treatment from the Hamiltonian in the helical base
\begin{equation}
 H= \sqrt{1+\lambda^2} \sum_{k,s} (\cos(k +s \phi(\lambda)) - \mu)n_{k,s} + H_U
\end{equation}
where $H_U$ is given by \eqref{eq:Hubbard_Interaction_helical}.
Decomposing the fermionic operators $c_s(x)$ into left- and right-movers we have
\begin{equation}
 c_s(x) = e^{i k^{L,s}_F x} c_{L,s}(x) + e^{i k_F^{R,s} x} c_{R, s}(x).
\end{equation}
Linearizing the non-interacting theory around the four Fermi points we find for the non-interacting 
part
\begin{equation}
H_0 = v_F(\lambda) \sum_{k,s} k n_{R,s}(k) - v_F(\lambda) \sum_{k,s} k n_{L,s}(k)
\label{eq:H0bosonized}
\end{equation}
where only the information about the Fermi velocity enters.
We like to remind the reader that since in \eqref{eq:vf} the phase-shift has dropped out, \eqref{eq:H0bosonized} contains no information about the position 
of the four Fermi-points.
To bosonize the interaction we note 
that for the particle-density we have
\begin{equation}
 \begin{split}
  n_s(x) & = c^\dagger_s(x) c_s(x) \\
  & = \sum \limits_{\alpha=R,L} c^\dagger_{\alpha ,s} c^{\phantom{\dagger}}_{\alpha,s} + c^\dagger_{\alpha, s} c^{\phantom{\dagger}}_{-\alpha, s} e^{-i2\alpha x k_F^0}
 \end{split}
\end{equation}
where the dependence on the phase-shift has also dropped out and $k_F^0$ as given in \eqref{eq:kf0} contains only information on the original Fermi velocity. This means that the Hubbard interaction 
stays form-invariant. For these reasons we find that Hamiltonian \eqref{eq:H_basic} still has the same bosonized low-energy description $H_B$ as the plain Hubbard-model:
\begin{equation}
 H_B = H_C + H_S
\end{equation}
with
\begin{equation}
\hspace{-0.3em}H_C \hspace{-0.3em}= 2 \pi v_F(\lambda) \hspace{-0.3em}\int \hspace{-0.3em}dx \hspace{-0.1em}\left[(1 + \frac{U}{v_F(\lambda)})(\partial_x \theta_C)^2 + (\partial_x \phi_C)^2\right]
\end{equation}
and% for the spin sector
\begin{equation}
\hspace{-0.3em}H_S \hspace{-0.3em}= 2 \pi v_F(\lambda) \hspace{-0.3em}\int \hspace{-0.3em}dx \hspace{-0.1em}\left[(1 - \frac{U}{v_F(\lambda)})(\partial_x \theta_S)^2 + (\partial_x \phi_S)^2\right].
\end{equation}
We have omitted the umklapp term in the charge sector and we have assumed that $U> 0$ so that the spin-sector acquires no mass-gap.
In terms of the usual left moving fields $\phi_{L,s}$ and right-moving fields $\phi_{R,s}$ of helicity $s$
we have introduced the fields
\begin{eqnarray}
 \phi_s &= \frac{1}{2\sqrt{\pi}}(\phi_{L,s} - \phi_{R,s}) \\
 \theta_s &= \frac{1}{2\sqrt{\pi}} (\phi_{L,s} + \phi_{R,s}).
\end{eqnarray}
For the finally used charge and spin fields we derive
\begin{eqnarray}
 X_C & = \frac{1}{\sqrt{2}} (X_+ + X_-) \\
 X_S & = \frac{1}{\sqrt{2}} (X_+ - X_-)
\end{eqnarray}
where $X$ is either the symbol $\theta$ or $\phi$.
We now proceed to show that this low-energy Hamiltonian satisfies the same symmetry as the original lattice Hamiltonian.
Simple algebra shows that 
\begin{equation}
 v_F(\lambda,\mu) = \nu_m(\lambda)v_F(\lambda_m, \frac{\mu}{\nu_m(\lambda)})
\end{equation}
and hence again this implies 
\begin{equation}
 H_B(\lambda, \mu, U) = \nu_m(\lambda) H_B(\lambda_m, \frac{\mu}{\nu_m(\lambda)}, \frac{U}{\nu_m(\lambda)}).
\end{equation}
From the decomposition into left and right-movers
\begin{equation}
 c_s(x) = e^{i (-k^0_F- s\phi) x} c_{Ls}(x) + e^{i (k_F^0 - s \phi) x} c_{R s}(x)
\end{equation}
we see that applying the same gauge transform separately to the left and right movers enables the removal of the phase-shift $\phi$.
Finalizing the derivation of the Hamiltonian we have
\begin{eqnarray}
  &H_B = \sum \limits_{a = C,S} H_a\\
\label{eq:bosonic_hamiltonian_summary}
  &H_a  = 2 \pi v_F(\lambda) K^2_a  \int dx \left[(\partial_x \overline{\theta}_a)^2 + (\partial_x \overline{\phi}_a)^2\right]
\end{eqnarray}
with
\begin{equation}
 K_{C/S} = \left( 1 \pm \frac{U}{v_F(\lambda)}\right)^{\frac{1}{4}},
\end{equation}
$\theta_{C/S} = \frac{1}{K_{C/S}} \overline{\theta}_{C/S}$ and $\phi_{C/S} = K_{C/S} \overline{\phi}_{C/S}$.
\subsubsection{Observables}
With the knowledge of the low-energy Hamiltonian \eqref{eq:bosonic_hamiltonian_summary} we can derive spin spin correlation functions.
We find
\begin{equation}
 S^{zz}(x,\tau) = \text{Re} \left( e^{i 2 \phi x} (A_S^\phi A_S^\theta + A_S^\phi A_C^\theta \cos{2 k_0 x} ) \right)
\label{eq:Szz_bosonized_final}
\end{equation}
with
\begin{equation}
\begin{split}
 A^\phi_S &= \frac{1}{a} \left(\frac{\beta_S}{\pi} \sin \left( \frac{\pi}{\beta_S } z \right) \right)^{-K_S^2} \\
 A^\theta_S &= \frac{1}{a} \left(\frac{\beta_S}{\pi} \sin \left( \frac{\pi}{\beta_S} z \right) \right)^{-\frac{1}{K_S^2}}\\
 A_C^\theta &= \frac{1}{a} \left(\frac{\beta_C}{\pi} \sin \left( \frac{\pi}{\beta_C} z \right) \right)^{-\frac{1}{K_C^2}}.
\end{split}
\end{equation}
Here $\beta_{C/S} = 2\pi \beta v_F(\lambda) K^2_{C/S}$ and
we have left the $z = \tau+i x + a$ dependence for the $A$'s implicit in the notation.
Looking at \autoref{fig:szsz} we see that the low-energy features are predicted correctly. We have peaks at $k=\pi \pm \phi$ and no peak at $k=\pi$.
Also we are consistent with the predictions of the Heisenberg-limit outlined in the next subsection.
For $S^\pm$ we find
\begin{equation}
\begin{split}
 &S^\pm(x,\tau) = S^{zz}(x,\tau) + \\
 %\langle B(z) B(0) \rangle 
 &+A_C^\theta A_S^\theta \cos(2k_F^0 x)
 - \frac{4\pi^2}{ K_S^2\beta_S^2} \csc^2\left( \frac{\pi}{\beta_S}z\right).
\end{split}
\end{equation}
This is consistent with the spectra in \autoref{fig:spsm}, especially the peak at $k=\pi$ is correctly predicted.
The location of the peaks at $k=\pi \pm \phi$ is identical to those in $S^{zz}$, a prediction that is also 
made by a consideration of the Heisenberg-limit. A very detailed analysis of the critical exponents from a Bethe Ansatz solution is given in Ref.~\cite{PhysRevB.86.085126}.
\subsection{The Heisenberg-limit}
The identity between the Hubbard and the Rashba-Hubbard model has implications for the strong-coupling limit.
Here we consider the half-filled case with non-interacting Hamiltonian
\begin{equation}
 H_0 = \sum \limits_r \vec{c}^{\hspace{0.1em}\dagger}_r T \vec{c}^{\phantom{\dagger}}_{r+1}
\end{equation}
with 
\begin{equation}
 T = t \mathbb{1} + i \lambda \sigma_y.
\end{equation}
\subsubsection{Derivation}
The transform to the helical base can be facilitated by \eqref{eq:fermionic_transformation}.
This gives
\begin{equation}
 H_0 = \sum \limits_r \gamma_r^\dagger (t + i \lambda \sigma_z) \gamma^{\phantom{\dagger}}_{r+1}
\end{equation}
with $\gamma_r$ denoting a spinor in the helical base.
We can insert the twist by using
\begin{equation}
 t + i \lambda \sigma_z = \sqrt{t^2 + \lambda^2} e^{i\phi \sigma_z}
\end{equation}
Therefore we find that $H_0$ has the 
form
\begin{equation}
 H_0 = \sqrt{t^2+\lambda^2} \sum \limits_r \eta^\dagger_r \eta^{\phantom{\dagger}}_{r+1}
\end{equation}
with fermions given by
\begin{equation}
 \eta_r^\dagger = \gamma_r^\dagger e^{-i\phi \sigma_z r}
\end{equation}
which has the isotropic Heisenberg model $H_{\text{iso}}$ as strong-coupling limit:
\begin{equation}
 H_{\text{iso}} = \frac{4(t^2 + \lambda^2)}{U} \sum \limits_r \vec{S}_r^\eta \vec{S}_{r+1}^\eta.
\label{eq:Hiso}
\end{equation}
This implies a representation of the spin-operator in terms of fermions as
\begin{equation}
 \vec{S}^\eta_r = \eta_r^\dagger \vec{\sigma} \eta^{\phantom{\dagger}}_r.
\end{equation}
Now we start to twist back \eqref{eq:Hiso} where we find 
\begin{equation}
\begin{split}
 \vec{S^\eta_r} &= c_r^\dagger S^\dagger e^{-i\phi \sigma_z r} \vec{\sigma} e^{i\phi \sigma_z r} S c^{\phantom{\dagger}}_r\\
 & = R(\frac{\pi}{2}, \vec{e_x}) R(2\phi r, \vec{e_z}) \vec{S}^c_r.
 \end{split} 
\end{equation}
Therefore we find for the isotropic Heisenberg model
\begin{equation}
 H_{\text{iso}} = \frac{4(t^2 + \lambda^2)}{U} \sum \limits_r \vec{S}_r^c R(2\phi, \vec{e_z})\vec{S}_{r+1}^c.
\end{equation}
Evaluating the rotation matrix we find that the following extended anisotropic Heisenberg-model corresponds to the isotropic Heisenberg model after the transform:
\begin{equation}
\begin{split}
H &= H_{\text{aniso}} + H_{\text{DM}}\\
H_{\text{aniso}} &= \frac{4}{U} \left(J_\parallel (S_r^x S_{r+1}^x + S_r^y S_{r+1}^y) + J_\perp S_r^z S_{r+1}^z \right)\\
H_{\text{DM}} &= - \frac{8t}{U} \lambda \left( S_r^x S_{r+1}^y - S_r^y S_{r+1}^x\right)
\end{split}
\end{equation}
with $J_\parallel = t^2 - \lambda^2$ and $J_\perp = t^2 + \lambda^2$.
This corresponds to an anisotropic Heisenberg-model with an added DM interaction that is pointing in the $S^z$ direction. It
is well-known\cite{PhysRevB.64.094425} that the DM interaction can be gauged away by a gauge transform on the $S^\pm$ operators.
\subsubsection{Observables}
\label{subsubsec:HeisenbergObservables}
The static spin spin correlations in the $\eta$-basis are
\begin{equation}
 \langle S_r^{\eta,\alpha} S_0^{\eta,\alpha} \rangle \propto (-1)^r \frac{\ln^{\frac{1}{2}}(r)}{r}
\end{equation}
in the long wavelength limit\cite{0305-4470-31-20-002}. This expression is valid for each spin component $\alpha$ since in the $\eta$-basis $SU(2)$ spin symmetry is present.
From this result one can obtain the spin spin correlations by twisting back into the $\uparrow \downarrow$-basis:
\begin{eqnarray}
\vec{S}^\eta_r = R(\frac{\pi}{2}, \vec{e}_x) R(2\phi r, \vec{e}_z) \vec{S}^{\uparrow \downarrow}_r.
\end{eqnarray}
With that we find for the correlation functions in the $\uparrow \downarrow$ basis
\begin{equation}
\begin{split}
&\langle S_\alpha^{\uparrow \downarrow}(r) S_\alpha^{\uparrow \downarrow}(0)\rangle = \\
&= \hspace{-0.2em}\left[R(\frac{\pi}{2}, \vec{e}_x) R(2\phi r, \vec{e}_z) R^T(\frac{\pi}{2}, \vec{e}_x)\hspace{-0.2em}\right]_{\alpha,\alpha} \hspace{-0.5em}\langle S^{\eta}_\alpha(r) S^{\eta}_\alpha(0) \rangle.
\end{split}
\end{equation}
Evaluating the matrix we find for the components:
\begin{eqnarray}
 \langle S_x^{\uparrow \downarrow}(r) S_x^{\uparrow \downarrow}(0)\rangle  \propto & \cos(2\phi r) (-1)^r \frac{\ln^{\frac{1}{2}}(r)}{r} \\
 \langle S_y^{\uparrow \downarrow}(r) S_y^{\uparrow \downarrow}(0)\rangle  \propto & (-1)^r \frac{\ln^{\frac{1}{2}}(r)}{r} \\
 \langle S_z^{\uparrow \downarrow}(r) S_z^{\uparrow \downarrow}(0)\rangle  \propto & \cos(2\phi r) (-1)^r \frac{\ln^{\frac{1}{2}}(r)}{r}.
\end{eqnarray}
\begin{figure}
\begin{tabular}{cc}
\subfloat[\strut
]{
 \includegraphics[width=0.47\linewidth]{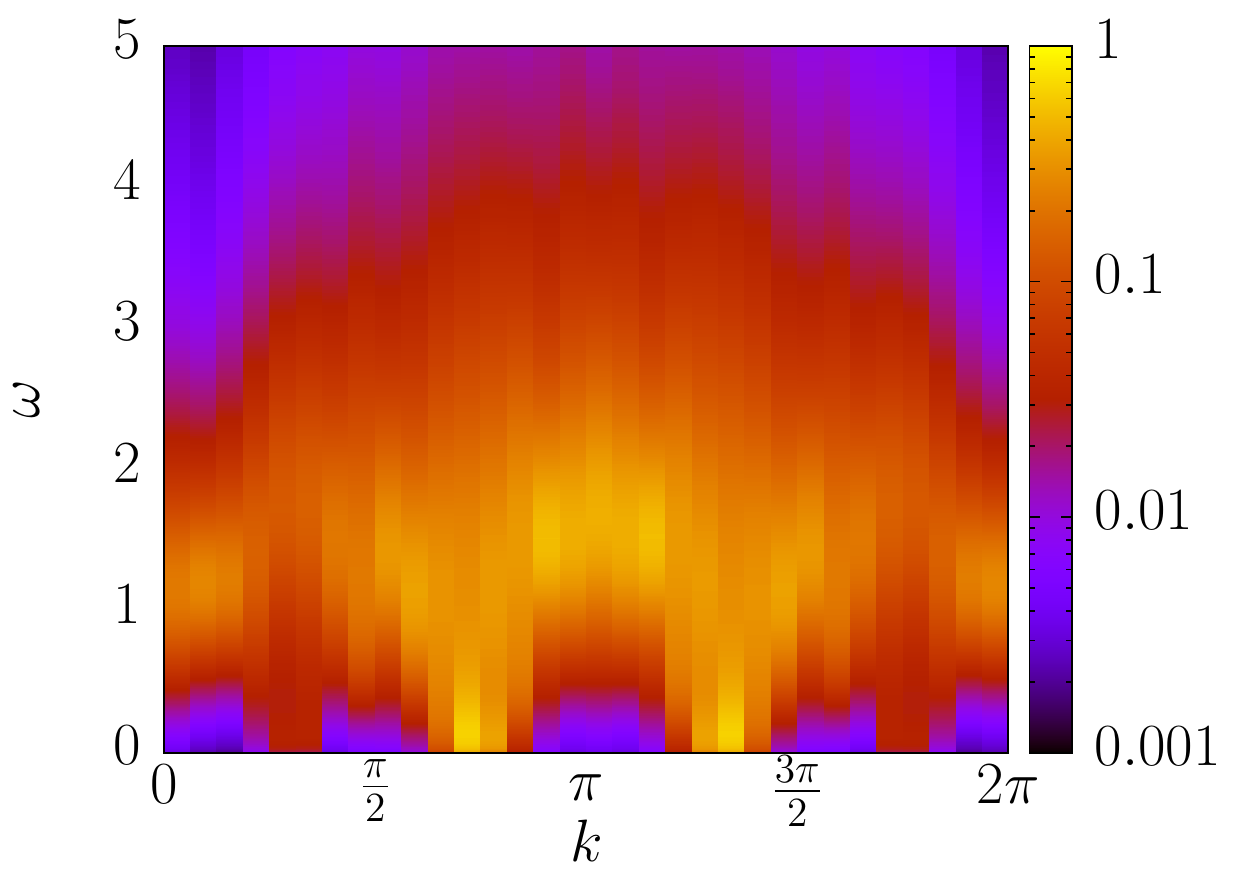}} &
\subfloat[\strut
]{
 \includegraphics[width=0.47\linewidth]{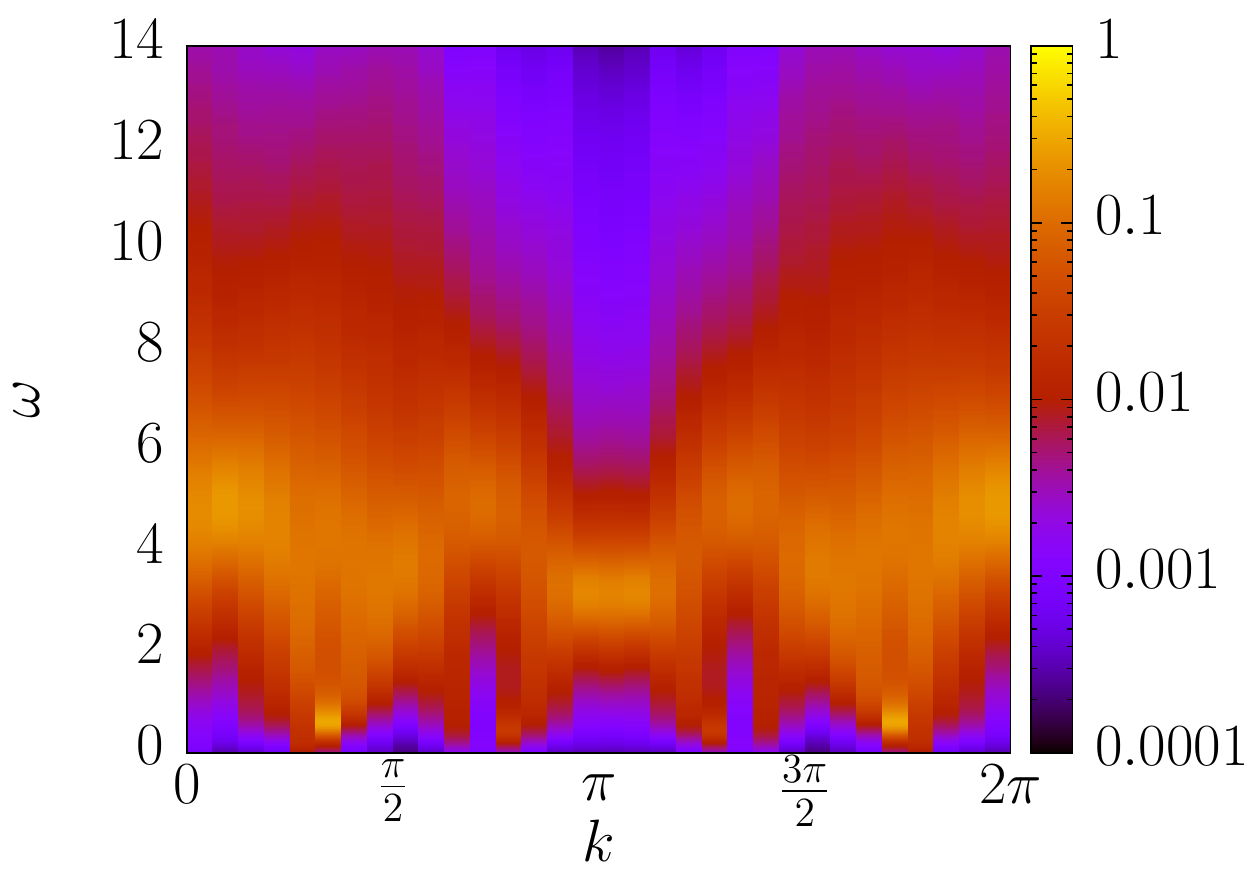}} \\
\end{tabular}
\caption{$\langle S^z S^z\rangle(k,\omega)$ correlation functions at $U=3$ and $\beta=10$ from a Monte-Carlo simulation.
(a) has $\lambda = 0.5$ which gives $\phi\approx 0.15\pi$ whereas (b) has $\lambda = 2$ with $\phi \approx 0.35\pi$.
}
\label{fig:szsz}
\end{figure}
This is consistent with the Monte-Carlo data of \autoref{fig:szsz} which was simulated using the CT-INT method \cite{Rubtsov2005}. We clearly see the two low energy peaks located symmetrically around $k=\pi$
Using the $x$ and $y$ components of the spin-vector we find
\begin{equation}
\begin{split}
 &\langle S_+^{\uparrow \downarrow}(r) S_-^{\uparrow \downarrow}(0)\rangle + \langle S_-^{\uparrow \downarrow}(r) S_+^{\uparrow \downarrow}(0)\rangle\\
 &\propto \cos(2\phi r)(-1)^r \frac{\ln^{\frac{1}{2}}(r)}{r} + (-1)^r \frac{\ln^{\frac{1}{2}}(r)}{r}
 \end{split}
\end{equation}
\begin{figure}
\begin{tabular}{cc}
\subfloat[\strut
]{
 \includegraphics[width=0.47\linewidth]{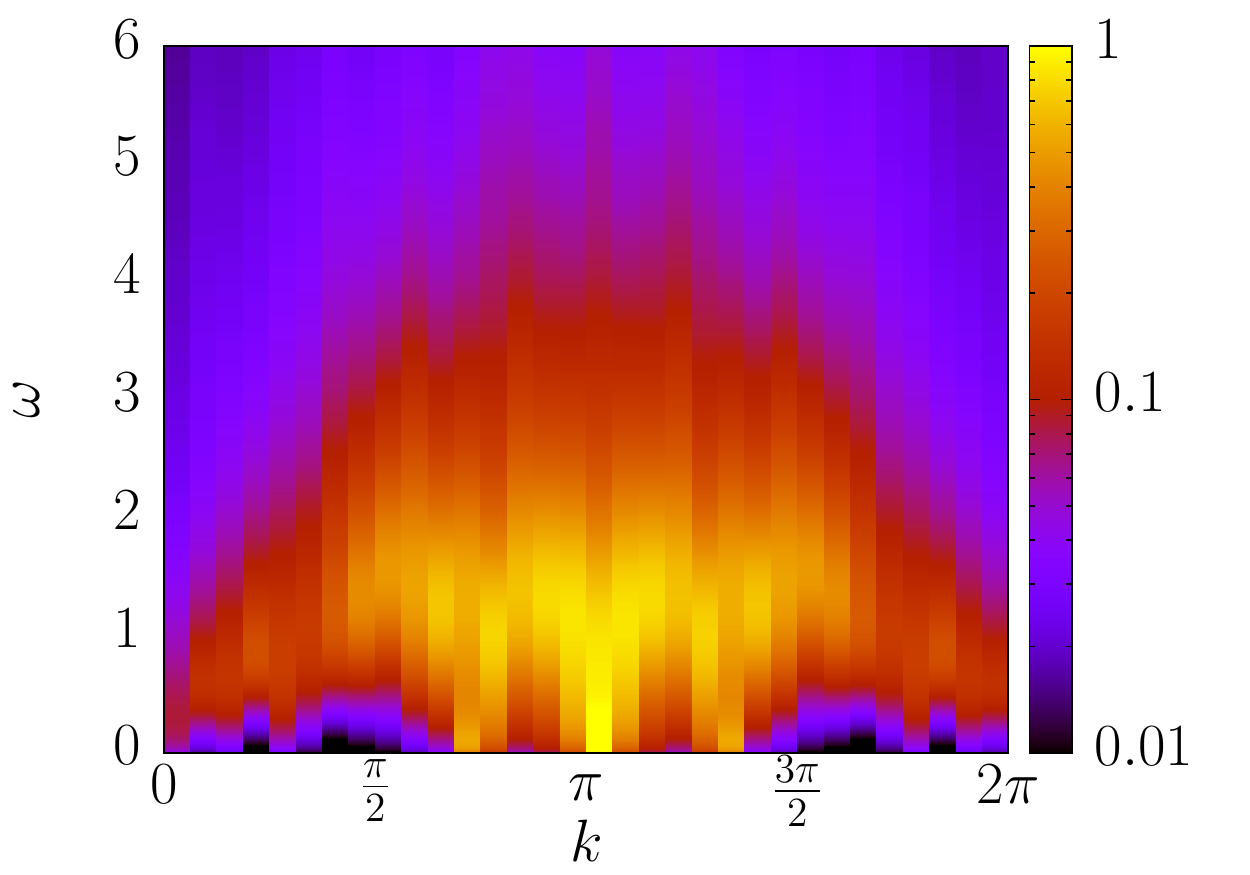}} &
\subfloat[\strut
%$S^{\pm}(k,\omega), \lambda = 2, U=3, \beta=10$
]{
 \includegraphics[width=0.47\linewidth]{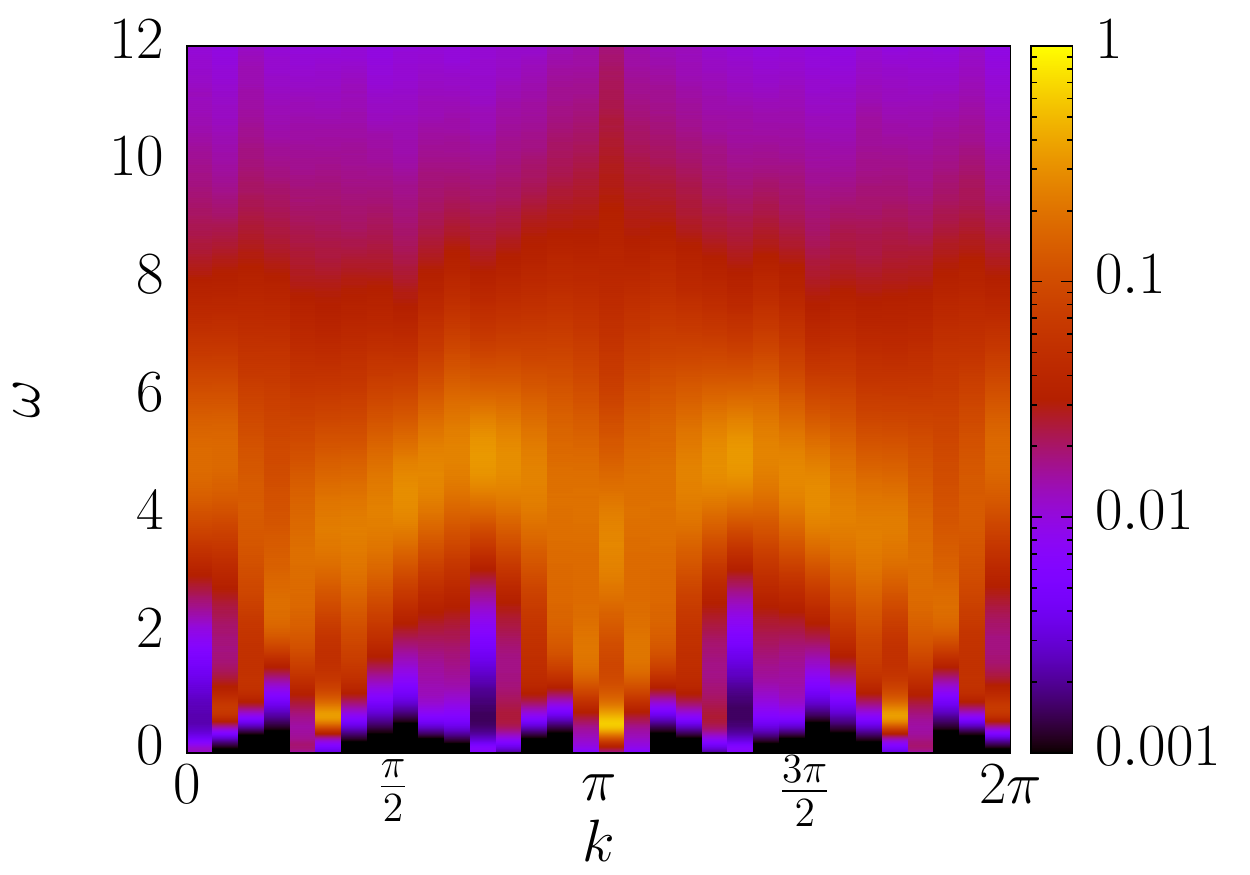}} \\
\end{tabular}
\caption{$\langle S^+ S^-\rangle(k, \omega)$ correlation functions at $U=3$ and $\beta=10$ from a Monte-Carlo simulation.
(a) has $\lambda = 0.5$ which gives $\phi\approx 0.15\pi$ whereas (b) has $\lambda = 2$ with $\phi \approx 0.35\pi$.}
\label{fig:spsm}
\end{figure}
We also see this in our CT-INT spectra in \autoref{fig:spsm}. We have one contribution pinned to $k=\pi$ and other contributions located symmetrically around $k=\pi$ identical to what is found in
$\langle S_z^{\uparrow \downarrow}(r) S_z^{\uparrow \downarrow}(0)\rangle$.
This mapping equally allows a reinterpretation of the full dynamical spin structure factors computed in \cite{2013NatPh9435M} for the isotropic Heisenberg model
to situations with Rashba spin orbit coupling.

\section{Generalizations}
\label{sec:generalizations}
\subsection{Long-Range interaction}
The mapping naturally lends itself to some generalizations.
First we have Hubbard models with a long-range Coulomb interaction.
\begin{equation}
 H_{LR} = H_0(\lambda) + \sum \limits_{r} V(r) \sum_i n_{i+r} n_i
\end{equation}
where $n_i = n_{i,\uparrow} + n_{i,\downarrow}$. Again, the spin orbit interaction can be rescaled into the coupling parameters:
\begin{equation}
 H_{LR}(\lambda) = \frac{1}{|c_{m}|} \left(H_0(0) + \sum \limits_{r} \tilde{V}(r) \sum_i n_{i+r} n_i\right)
\end{equation}
with $\tilde{V}(r) = V(r) |c_{m}|$.
\subsection{Coupling to the spin}
This can be further extended to anisotropic Hubbard-models by adding spin-terms.
In particular we consider the additional term
\begin{equation}
 H^{yy} = \sum \limits_{r} S^y_{r+1} S^y_r
\end{equation}
with $S^y_r = \frac{-i}{2}(c^\dagger_{r,\uparrow} c^{\phantom{\dagger}}_{r,\downarrow} - c^\dagger_{r,\downarrow} c^{\phantom{\dagger}}_{r,\uparrow})$.
Performing the transform to helical electrons we find that $S^y_r$ given in physical spins transforms to $- S^z_r = \frac{1}{2} (n_{r,-} - n_{r,+})$,
given in terms of helical spins,
which is manifestly invariant under the gauge-transform.
This invariance of $S^z$ can be used to additionally include an in-plane magnetic field with coupling strength $b$ in $y$-direction.
\begin{equation}
 H_{\text{mag}} = b \sum \limits_r S^y_r
\end{equation}
which transforms to 
\begin{equation}
 H_{\text{mag}} = -b \sum \limits_r S^z_r
\end{equation}
and is again invariant under the gauge-transform.
\subsection{Phonons}
Our results can be further generalized to electron-phonon models with
Holstein type electron-phonon coupling\cite{2013PhRvB..87g5149H}.
Since the part of the Hamiltonian that couples electrons and phonons is given by
\begin{equation}
 H_{ep} = g\sum_i Q_i (n_i - 1)
\end{equation}
where $n_i = n_{i,\uparrow} + n_{i,\downarrow}$, we see that in this case the
transformation required to eliminate the Rashba term corresponds to a 
rescaling of the coupling strength $g$ with the bandwidth.
Further generalization to long-range electron-phonon interaction is possible. If the interaction is
\begin{equation}
 H_{ep} = \sum_{i,j} f_{i,j} Q_j \sum_\sigma \alpha_\sigma (n_{i,\sigma} - \frac{1}{2})
\end{equation}
and we assume a spin-independent $\alpha$ we find
\begin{equation}
 H_{ep} = \alpha \sum_{i,j} f_{i,j} Q_j (n_{i} - 1)
\end{equation}
which is again invariant under the transform to helical spins and the gauge transform.
\subsection{Disorder}
Potential disorder
\begin{equation}
 H_{\text{dis}} = \sum \limits_i \mu_i n_i
\end{equation}
couples only to the local particle density $n_i$ which is equally invariant
under the transformation to the helical basis and is not modified by the gauge transform.
Ref.~\onlinecite{Kaplan1983} discusses the case how to link different realizations of bond disorder.

\subsection{Long-Range hopping}
Only a very slight generalization is possible to include long-range hoppings.
Given a hopping Hamiltonian that includes long-range hopping with distances $d$,
\begin{equation}
 H_0 = \sum \limits_{d=1} H_d(t_d, \lambda_d)
\end{equation}
where for each $d$ we have
\begin{equation}
 H_d(t_d, \lambda_d) = \sqrt{t_d^2 + \lambda_d^2} \sum \limits_{k,s} \cos(k d+ s \phi_d) n_{k,s}
\end{equation}
with $\phi_d = \arctan(\frac{\lambda_d}{t_d})$. $H_d$ is already in the helical base which is always doable for the Rashba spin orbit interaction 
in one dimension.
We can only employ the $U(1)$ gauge symmetry globally if we require
\begin{equation}
 \phi_d = \frac{1}{d} \phi_1,
\end{equation}
where $\phi_1 = \frac{2 j \pi}{L}$, a value commensurate with the lattice. This enforces a 
particular form for the $\lambda_d$'s, namely
\begin{equation}
 \lambda_d = t_d \tan\left(\frac{2 j \pi}{d L} \right).
\end{equation}
With that we find for the hopping Hamiltonian
\begin{equation}
 \sum \limits_d H_d(t_d, \lambda_d) = \sum \limits_d H_d(\tilde{t_d},0)
\end{equation}
with
\begin{equation}
 \tilde{t_d} = \left| \frac{t_d}{\cos\left(\frac{2\pi j}{d L}\right)} \right|.
\end{equation}

\section{Summary and Conclusion}
\label{sec:summary}
We have carefully reviewed the mapping of the Rashba Hubbard model to the Hubbard model.
The validity of the mapping  for a material can in principle be tested by spin and angle resolved photoemission spectroscopy.
In one spatial dimension this implies that the transformation to the helicity basis is a global $SU(2)$ transformation (i.e. site or momentum independent).
Assuming global $SU(2)$ invariant interactions, such as long-range Coulomb interactions, coupling to the lattice and potential disorder, the helicity
is a good quantum number.
The mapping onto the $SU(2)$ symmetric Hubbard model requires a helicity dependent twist,
which places constraints on the form of the hopping matrix elements, and requires
commensurability between the lattice size and the value of the Rashba spin orbit coupling.
For spin resolved ARPES experiments and as explicitly shown for the Hubbard model,
this has for consequence that spin resolved spectra can be decomposed into two Hubbard type spinon-holon spectra, 
generically with different weights due to the device's spin quantization axis.
Both spectra map exactly onto each other when shifting the momentum in opposite directions.
For open boundary conditions, commensurability issues do not occur. 
A direct consequence of this mapping is to prove that it is justified to analyze the local density of states, Eq. (45),
in the realm of Luttinger liquid theory for the plain vanilla Hubbard model \cite{Claessen2011}.
At the two-particle level the result allows to understand the spin dynamics of the Mott insulating state of the
Rashba-Hubbard chain based on the results of the plain isotropic Heisenberg model\cite{2013NatPh9435M}.
The mapping equally impacts numerical simulations.
It enables one to reinterpret simulations of the Hubbard model in the Rashba-Hubbard setting.
In the present article we have shown this explicitly for  the spin resolved single particle spectral
function as well as for spin dynamics at half-band filling.
It is also interesting to point out that quantum Monte Carlo CT-INT simulations of the Rashba-Hubbard model,
are plagued by the negative sign problem. Hence, the mapping shows how to carry out a basis transformation to eliminate it.

Generalizations of this mapping to higher dimensions with larger coordination require fine tuning
by choosing parameters where  the spin orbit interaction remains effectively 
one dimensional\cite{PhysRevLett.90.146801,PhysRevLett.97.236601}, as already shown by Kaplan\cite{Kaplan1983}.
In light of the very special and robust features encountered in one dimensional
chains with Rashba spin orbit interactions, it is certainly very interesting to revisit the dimensional crossover\cite{PhysRevLett.87.276405,PhysRevB.56.7232, PhysRevLett.109.126404, PhysRevB.88.085120}.
In the one dimensional limit the mapping implies an $SU(2)$ symmetry\cite{PhysRevLett.97.236601} which will 
generically break down in higher dimensions or when chains are coupled to form ladder systems\cite{PhysRevB.88.045102,PhysRevB.83.115301}.
It is further expected that in this crossover regime the interplay between low-dimensionality and spin orbit coupling may lead to realizations
of the Fulde-Ferrell-Larkin-Ovchinnikov type\cite{PhysRevB.88.205127,PhysRevA.88.023622,2013NatCo4E2711Z} superfluidity.

\begin{acknowledgments}
We acknowledge fruitful discussions with J.~Aulbach, R.~Claessen and J.~Schäfer.
We thank Hosho Katsura, Evgeny Sherman and Arun Paramekanti for further references.
FG would like to thank D.~Luitz for reading early drafts of the manuscript.
FG acknowledges support from DFG Grant No.~AS120/9-1
and FFA from Grant No AS120/6-2 (FOR1162).
We thank the J\"{u}lich Supercomputing
Centre for generous allocation of CPU time.
The authors gratefully acknowledge the Gauss Centre for Supercomputing e.V. (\url{www.gauss-centre.eu})
for funding this project by providing computing time on the GCS Supercomputer SuperMUC at Leibniz Supercomputing Centre (LRZ, \url{www.lrz.de}).
\end{acknowledgments}
\bibliography{references}
\end{document}